\documentclass[sigconf]{acmart}

\AtBeginDocument{%
  \providecommand\BibTeX{{%
    \normalfont B\kern-0.5em{\scshape i\kern-0.25em b}\kern-0.8em\TeX}}}

\copyrightyear{2026}
\acmYear{2026}
\setcopyright{cc}
\setcctype{by}
\acmConference[CHI '26]{Proceedings of the 2026 CHI Conference on Human Factors in Computing Systems}{April 13--17, 2026}{Barcelona, Spain}
\acmBooktitle{Proceedings of the 2026 CHI Conference on Human Factors in Computing Systems (CHI '26), April 13--17, 2026, Barcelona, Spain}
\acmPrice{}
\acmDOI{10.1145/3772318.3790537}
\acmISBN{979-8-4007-2278-3/2026/04}

\usepackage{caption}
\usepackage{subcaption}

\graphicspath{ {figures/} }

\usepackage{color}
\usepackage{soul}    
\usepackage{gensymb} 

\usepackage{enumitem} 

\newcommand{\refine}[1]{\textcolor{black}{#1}}

\usepackage{listings}
\usepackage{caption}

\lstdefinestyle{promptstyle}{
    basicstyle=\ttfamily\small,
    breaklines=true,                   
    breakatwhitespace=true,            
    columns=fullflexible,              
    keepspaces=true,                   
    showstringspaces=false,            
    upquote=true,                      
    frame=single,
    framerule=0pt,
    backgroundcolor=\color{gray!10},
    numbers=none,
    xleftmargin=0em,
    xrightmargin=0em,
}

\lstdefinestyle{promptstyleAppendix}{
    basicstyle=\ttfamily\footnotesize,
    breaklines=true,
    breakatwhitespace=true,
    columns=fullflexible,
    keepspaces=true,
    showstringspaces=false,
    frame=single,
    framerule=0pt,
    backgroundcolor=\color{gray!10},
    numbers=none,
    xleftmargin=0em,
    xrightmargin=0em,
}

\usepackage{xspace} 

\copyrightyear{2026}
\acmYear{2026}
\setcopyright{cc}
\setcctype{by}
\acmConference[CHI '26]{Proceedings of the 2026 CHI Conference on Human Factors in Computing Systems}{April 13--17, 2026}{Barcelona, Spain}
\acmBooktitle{Proceedings of the 2026 CHI Conference on Human Factors in Computing Systems (CHI '26), April 13--17, 2026, Barcelona, Spain}
\acmPrice{}
\acmDOI{10.1145/3772318.3790537}
\acmISBN{979-8-4007-2278-3/2026/04}

\begin{document}

\title[VLM Persona vs Real Human]{Peeking Ahead of the Field Study: Exploring VLM Personas as Support Tools for Embodied Studies in HCI}

\author{Xinyue Gui}
\orcid{0000-0001-6541-224X}
\affiliation{%
  \institution{The University of Tokyo}
  \city{Tokyo}
  \country{Japan}
}
\email{xinyueguikwei@gmail.com}

\author{Ding Xia}
\orcid{0000-0002-4800-1112}
\affiliation{%
  \institution{The University of Tokyo}
  \city{Tokyo}
  \country{Japan}}
\email{dingxia1995@gmail.com}

\author{Mark Colley}
\orcid{0000-0001-5207-5029}
\affiliation{%
  \institution{UCL Interaction Centre}
  \city{London}
  \country{United Kingdom}
}
\email{m.colley@ucl.ac.uk}

\author{Yuan Li}
\orcid{0009-0005-8013-0896}
\affiliation{%
  \institution{The University of Tokyo}
  \state{Tokyo}
  \country{Japan}
}
\email{liyuan@keio.jp}

\author{Vishal Chauhan}
\orcid{0009-0003-4901-9917}
\affiliation{%
  \institution{The University of Tokyo}
  \city{Tokyo}
  \country{Japan}
}
\email{vishalchauhan@g.ecc.u-tokyo.ac.jp}

\author{Anubhav Anubhav}
\orcid{0000-0002-2480-6119}
\affiliation{%
  \institution{The University of Tokyo}
  \city{Tokyo}
  \country{Japan}
}
\email{anubhav2901@g.ecc.u-tokyo.ac.jp}

\author{Zhongyi Zhou}
\orcid{0000-0002-1363-7313}
\affiliation{%
  \institution{Google}
  \city{Tokyo}
  \country{Japan}
}
\email{zhongyi.zhou.work@gmail.com}

\author{Ehsan Javanmardi}
\orcid{0000-0003-0337-115X}
\affiliation{%
  \institution{The University of Tokyo}
  \city{Tokyo}
  \country{Japan}
}
\email{ejavanmardi@g.ecc.u-tokyo.ac.jp}

\author{Stela Hanbyeol Seo}
\orcid{0000-0002-0412-476X}
\affiliation{%
  \institution{Kyoto University}
  \city{Kyoto}
  \country{Japan}
}
\email{stela.seo@i.kyoto-u.ac.jp}

\author{Chia-Ming Chang}
\orcid{0000-0003-0390-6361}
\affiliation{%
  \institution{National Taiwan University of Arts}
  \city{Taipei}
  \country{Taiwan}
}
\email{info@chiamingchang.com}

\author{Manabu Tsukada}
\orcid{0000-0001-8045-3939}
\affiliation{%
  \institution{The University of Tokyo}
  \city{Tokyo}
  \country{Japan}
}
\email{mtsukada@g.ecc.u-tokyo.ac.jp}

\author{Takeo Igarashi}
\orcid{0000-0002-5495-6441}
\affiliation{%
  \institution{The University of Tokyo}
  \city{Tokyo}
  \country{Japan}
}
\email{takeo@acm.org}


\begin{abstract}
Field studies are irreplaceable but costly, time-consuming, and error-prone, which need careful preparation. Inspired by rapid-prototyping in manufacturing, we propose a fast, low-cost evaluation method using Vision-Language Model (VLM) personas to simulate outcomes comparable to field results. While LLMs show human-like reasoning and language capabilities, autonomous vehicle (AV)-pedestrian interaction requires spatial awareness, emotional empathy, and behavioral generation. This raises our research question: To what extent can VLM personas mimic human responses in field studies? We conducted parallel studies: 1) one real-world study with 20 participants, and 2) one video-study using 20 VLM personas, both on a street-crossing task. We compared their responses and interviewed five HCI researchers on potential applications. Results show that VLM personas mimic human response patterns (e.g., average crossing times of 5.25 s vs. 5.07 s) lack the behavioral variability and depth. They show promise for formative studies, field study preparation, and human data augmentation.
\end{abstract}



\begin{teaserfigure} 
  \includegraphics[width=\textwidth]{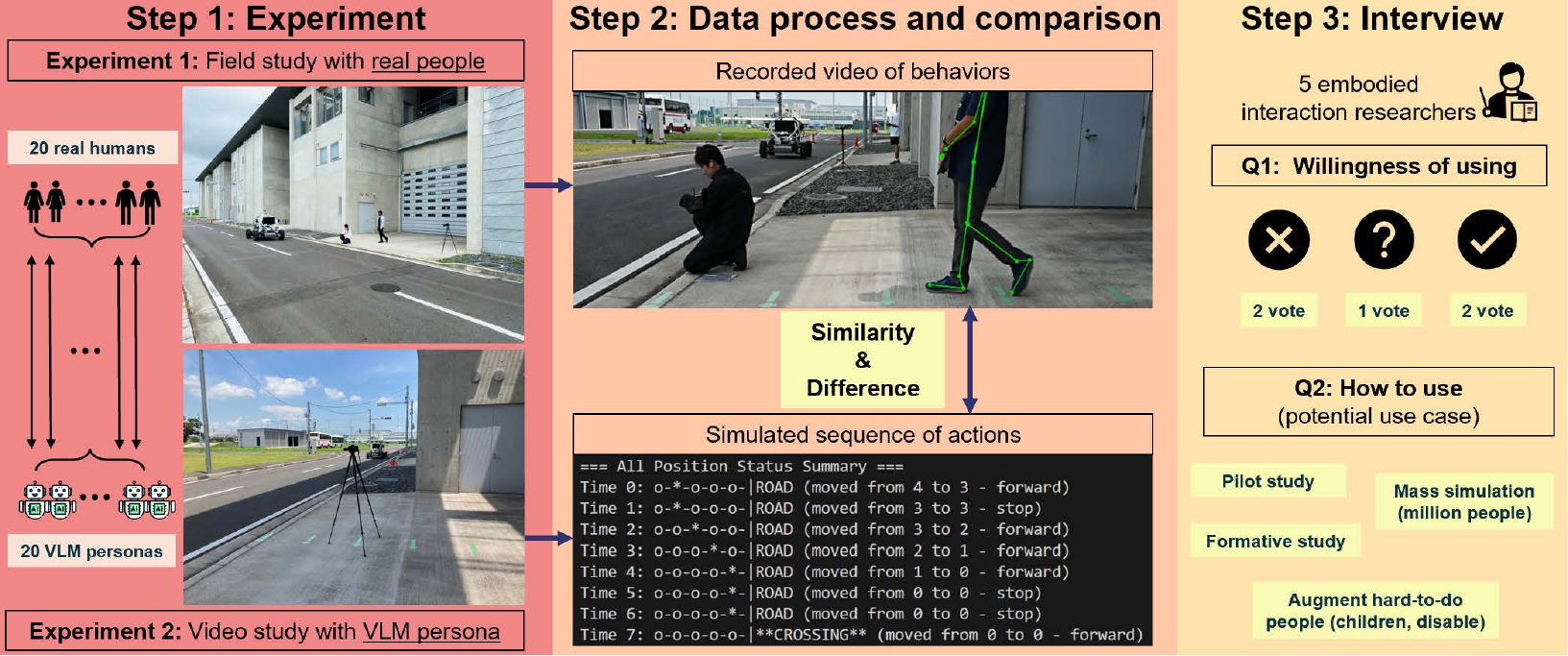}
  \caption{Overview of the experimental process. Step 1 (red): Two parallel studies were conducted: 1) a field study with 20 human participants and 2) a video-based study with 20 VLM personas modeled after them. Step 2 (peach): Data were preprocessed into consistent formats for comparison, followed by analysis of similarities and differences. Step 3 (yellow): Interviews were conducted with five HCI researchers to gather feedback on the use of VLM personas in research.}
  \Description{Overview of our experimental process. Step 1 (red area): We conducted two parallel studies 1) a field study with 20 real participants and 2) a video-based study using 20 VLM personas modeled after them. Step 2 (peach area): We pre-processed the data to ensure consistent formats for comparison, analyzing behavioral and emotional similarities and differences. Step 3 (yellow area): We shared the simulation and comparison results with five HCI researchers who conduct field studies and interviewed them to gather feedback on using VLM personas in research.}
  \label{fig:teaser}
\end{teaserfigure}

\maketitle
\renewcommand{\shortauthors}{Gui et al.}

\section{Introduction}  

Evaluating the interaction between autonomous vehicles (AVs) and pedestrians is essential for the safe and effective deployment of AVs. Understanding how pedestrians perceive and react to AVs informs both algorithm development and human-centric interaction design \cite{mahadevan2019av}. In the context of studying AV-pedestrians' interaction, field studies are important \cite{clarke2024llm} because they capture natural human responses (e.g., spontaneous behavior and emotion), which are difficult to replicate in lab environments \cite{sirkin2016embodied, gerig2018missing}. However, conducting field studies presents methodological challenges. Controlled environments limit behavioral authenticity, while real-world testing raises concerns about safety and ethics \cite{geelhoed2024discomfort, reig2018field}. Participants may misjudge situations, which can lead to accidents or increased stress. In addition, field studies require AV hardware, dedicated space, multiple experimenters, and extensive pilot testing \cite{10.1145/2809730.2809755}, which highlights the importance of careful study design beforehand.

To address these challenges, researchers explored alternative approaches that approximate real-world responses. One direction uses mixed reality (MR) and tangible interfaces to enhance immersion without physical risk \cite{10.1145/3411764.3445159, gao2025enhancing, dufresne2024touching}. However, controlled settings restrict natural movement, which is problematic for observational studies requiring spontaneous behavior \cite{gerig2018missing}. Extended Reality (XR) methods also lack a strong sense of presence due to technical limitations, especially the trust judgments \cite{abraham2022implications}. Another direction is to use AI agents to simulate human-like reasoning and responses \cite{liu2024make, wang2025can}. We enhance this approach by proposing to use Vision-Language Model (VLM) personas to simulate subjective and behavioral reactions to approaching AVs. In this work, we explore how VLM personas can support field studies. For example, VLMs can help inform pilot tests to reduce costly errors or augment human participant data to balance cost and analytical depth. These may provide a low-cost, human-like preview of real field study results, helping to reduce errors and allowing researchers to make early adjustments to prevent additional costs.

While VLMs show human-level performance in language tasks (e.g., product reviews or consultations) \cite{hung2024simtube, meng2024mmlscu, li2024llms}, performing field study tasks needs geospatial awareness \cite{gui2024text+}, emotional synthesis \cite{wang2024physiological}, and behavior generation \cite{rasouli2017understanding}. Although previous works (e.g., VLMs \cite{bordes2024introduction, zhou2024vision, shao2025large}) suggest that AI personas can now interpret spatial contexts and generate corresponding actions. However, it remains unclear how well their responses align with human results, both quantitatively and qualitatively. That is, \textbf{rather than relying solely on performance benchmarks, validating their utility requires a deeper comparison between AI-generated and human responses}. It also remains unclear how HCI researchers assess the value of using VLM personas in research, and what roles they envision for this tool. Therefore, our research questions were:

\smallskip

\noindent
\textbf{1) To what extent can VLM personas mimic human responses in field studies?}\newline
\textbf{2) How do HCI researchers whose work requires field studies envision their use in embodied research?}

\smallskip

To answer these questions, we conducted a three-step process (see \autoref{fig:teaser}). First, we ran two parallel experiments: 1) a field study with 20 real participants, and 2) a video-based simulation using VLM personas modeled from the same participants. We rented a real AV, reserved a large testing area, and installed a custom external human-machine interface (eHMI) \cite{carmona2021ehmi, bazilinskyy2021should} to enable pedestrian-AV interaction. Both the real participants and their VLM counterparts performed the same task (i.e., crossing the street in the presence of an AV). We applied the same metrics to both testing groups (real humans and VLM personas), analyzing behavioral data (e.g., discretized trajectories, crossing times) and subjective data (e.g., 5-point Likert ratings on confidence, trust, and textual responses from post-study interviews).

Results indicate that the behavioral patterns generated by VLM personas fall within the distribution of human responses. Trust ratings were comparable, though VLMs consistently reported overconfidence. Thematic analysis found that VLMs captured key human insights but lacked nuance in nonverbal cues and complex social behaviors. To address the second question, we interviewed five HCI researchers experienced in field studies. After presenting our methods and findings, we discussed the applicability of VLM personas in the researchers’ workflows. They recognized the benefits of using VLMs for: 1) conducting exploratory studies on unfamiliar topics to establish a research direction, 2) pilot testing and refining experimental designs, and 3) augmenting human data by simulating responses from hard-to-reach or vulnerable populations (e.g., children, the elderly, or individuals with disabilities). These can generate human-like data at low cost throughout the research process, reducing the risk of errors in subsequent formal studies. \refine{We extend our findings to broader embodied interaction, defining “embodied” as simulated personas rather than body-integrated technologies. An input–output analysis of embodiment literature shows how VLM persona simulations can generalize to other domains.}

\medskip

\noindent\fcolorbox{orange}{orange!30}{\textbf{Contribution Statement}~( \citet{Wobbrock.2016})}

\smallskip

\begin{itemize}[leftmargin=0.5cm]
\item \textbf{Method.} We introduce the use of VLM personas to simulate complex human behaviors and spatial perception in AV-pedestrian interactions, extending beyond language-based evaluations. 
\item \textbf{Empirical study.} We present an empirical study exploring how closely VLM personas align with real human responses quantitatively and qualitatively, and propose three guidelines for using VLM personas to better reflect human responses.
\item \textbf{Empirical study that tells us about how people use a system.} We investigate how 5 HCI researchers envision the role of VLM personas in embodied research, and offer a set of use-case guidelines based on their insights.
\end{itemize}

\section{Related Work}  

\subsection{Pedestrian analysis in human-AV interaction}

\refine{The shift from human-driven to self-driving vehicles introduces communication challenges \cite{rad2020pedestrians}. Pedestrians lose direct cues from drivers, such as eye contact and hand gestures, leading to increased uncertainty during interactions. That is, the AVs become “black boxes,” unable to clearly convey their perceptions or intentions \cite{carmona2021ehmi}. As a result, understanding and designing effective communication between pedestrians and autonomous vehicles has become a critical research topic \cite{10.1145/3699778}.} The evaluation of AV-pedestrian interaction typically contains objective and subjective metrics.

Objective metrics track whether participants perform actions that ensure their safety and achieve their goals (e.g., ``crossing the street,'' ``avoiding collision''). Research \cite{rasouli2017understanding} on crossing behavior shows that decisions and timing depend on numerous factors, including social factors (social norms, pedestrian flows), physical contexts (pedestrian state, traffic characteristics) \cite{colley2025improving}, and demographics (age, gender) \cite{hollander2021taxonomy, colley2021investigating, gehrke2023observed}. Subjective metrics capture a participant's internal state and perception. This is commonly measured using metrics like Trust in Automation (TiA) \cite{korber2018theoretical}, confidence \cite{wright2019agent}, attitudes \cite{nomura2006measurement}, or mental workload (e.g., NASA-TLX) \cite{hart1988development}.

A comprehensive analysis often combines both types of feedback. Sometimes, the subjective answer is coherent with and supplements the objective answer \cite{chauhan2023fostering}, but sometimes it is contradictory, showing the complexity of human perception and performance. For example, a participant might make a wrong decision (e.g., colliding with the vehicle in a video-based study) but report high confidence \cite{frank2019human, de2003effects}. Alternatively, a participant may perform a safe behavior when facing the AV but still feel a low sense of safety \cite{fink2025does, rasouli2019autonomous}.

These evaluation results show that the absence of a human driver creates a communication gap between the AV and the pedestrian \cite{rasouli2019autonomous}. To enhance mutual understanding, researchers introduced external human-machine interfaces (eHMIs) \cite{tran2024exploring}, which place the communicative interface on the exterior of the vehicle. Various eHMI modalities have been tested, including windshield text (e.g., ``Stop'') \cite{eisma2021external}, flashing light strips \cite{dey2020color, bazilinskyy2020external}, and projected virtual crosswalks \cite{bazilinskyy2019survey, ochiai2011homunculus}. Results show eHMIs can lead to more correct or faster crossing decisions, with text-based eHMIs often being the most straightforward \cite{bazilinskyy2019survey, chang2018video}. In addition, some researchers explore anthropomorphic \cite{lim2023anthropomorphic, bazilinskyy2019survey} and bionic eHMIs \cite{oudshoorn2021bio, wang2021virtual} to improve the noticeability and perception of devices and design intuitive interaction compared to standard eHMIs perceived as cold and neutral \cite{weidel2025enhancing}.

Anthropomorphic eHMIs mimic human-like cues, such as robotic eyes to simulate gaze \cite{chang2022can, wang2023my}, animated hands for gestures \cite{gui2024shrinkable, mahadevan2018communicating, epke2021see}, or facial expressions (e.g., smiling face) \cite{large2025face, smile}. Past research reveals significant individual differences in their reception. Some works report that these eHMIs improve the sense of safety \cite{bonneviot2023improve, li2025tailcue, jansen2023autovis}, providing a warmer atmosphere \cite{gui2023field, wang2023my}. Conversely, other works report feelings of discomfort or even being spied on \cite{gui2022going}. \refine{Several studies focusing on trust in anthropomorphic eHMIs have reported that anthropomorphic designs do not significantly increase trust in autonomous vehicles \cite{arabian2025systematic}. In addition, a field study found that eye-based eHMIs were not considered a primary factor influencing pedestrians’ trust \cite{gui2023field}.} As anthropomorphic cues rely on ``common sense'' norms, which differ between individuals \cite{de2022external}, their communicative efficiency can be highly polarized, with people tending to either strongly favor or strongly dislike it \cite{lim2023anthropomorphic, gui2024text+, large2023av}.

Therefore, comparing how anthropomorphic and non-anthro\-pomorphic eHMIs influence pedestrians' objective behavior (e.g., crossing) and subjective state (e.g., trust, psychological comfort) remains a central and unresolved topic \cite{bazilinskyy2019survey}. Thus, in our study, we selected one anthropomorphic and one non-anthropomorphic eHMI to explore whether humans and VLM personas exhibit similar behavioral tendencies, and whether the VLM personas can capture the complexity and individual differences observed in human responses.

\subsection{Evaluation Environment in AV-pedestrian Interaction} 

Field studies are widely used for their ability to capture pedestrians’ natural reactions to real vehicles. That is something video or virtual reality (VR)-based environments cannot fully replicate~\cite{10.1145/3699778, 10.1145/3290605.3300737}. \citet{sirkin2016embodied} identified three main types of field studies: Wizard-of-Oz setups \cite{3woz}, real-road AV simulators \cite{intent17}, and public deployment prototypes (e.g., the \textit{Ghost driver} platform)~\cite{10.1109/ROMAN.2016.7745210}. However, real-world evaluations require a highly controlled, labor-intensive setup \cite{35, 36, mahadevan2018communicating}. These studies often involve multiple on-site observers, extended testing periods, and tightly managed conditions \cite{gui2023field, yu2025peek}. In some cases, research teams rely on support from automotive companies, fabrication specialists, media crews, and collaborators across institutions \cite{44, 10.1145/2809730.2809755}. 

The challenges lie not only in the execution of the formal study but also in the extensive and costly preparation needed, such as pre-tests, rehearsals, and iterative refinements. Because a full-scale run is both time-consuming and costly, any issues that occur during the formal experiment have a disproportionately large impact, potentially undermining substantial portions of the collected data. Given these difficulties, it becomes essential to explore supportive tools. In our research, we tested VLM personas simulation, which can help reduce risk, improve preparation, and make field studies more efficient and robust.

\subsection{AI-generated Persona as an Evaluator in HCI Research} 

A persona is defined as a fictional, detailed user model that represents archetypal users \cite{blomkvist2002user}. Later developments in the persona lifecycle \cite{miaskiewicz2011personas, adlin2010essential, pruitt2010persona} showed that personas enhance the design process by giving designers a clear and evocative representation of the target user group. With the rise of LLMs, persona creation has become more accessible \cite{mcginn2008data, salminen2021survey}. Users can now generate virtual agents, such as personas defined by behavioral profiles that simulate human-like interaction, simply by writing prompts or integrating personality traits \cite{anvari2015effectiveness}, reducing the cost and effort of personification. This shift has led to a boost in the use of AI-generated personas in HCI research \cite{liu2024personaflow, prpa2024challenges}.

In HCI, LLM personas currently serve two main roles, each offering promise but also raising critical concerns. First is enabling AI agents to adopt specific roles when collaborating with users. For example, interacting as a user’s ``future self'' to promote personal growth \cite{jeon2025letters, khot2025challenging}, or acting as co-creative partners in educational and productivity contexts \cite{chu2025llm, qin2025synergy}. These agents can simulate inner dialogues \cite{liu2025proactive} and personalize interaction to boost engagement. The second role involves using LLM personas as evaluation tools to simulate user feedback. For example, reviewing videos \cite{hung2024simtube}, analyzing the impact of demographics \cite{ilkka2025demographics}, or participating in synthetic focus groups \cite{olawade2025enhancing}. Previous work has demonstrated that LLM personas can generate outputs that resemble human language and reasoning. However, relying too heavily on simulated feedback may give a false sense of validity. There is a risk to masking real human diversity and introducing biases from the LLM itself \cite{ye2024justiceprejudicequantifyingbiases}. Therefore, the results from LLM simulation always need to be critically examined.

These improvements show that the LLMs are acquiring the abilities of environmental perception and behavioral response. This in turn can provide field study researchers with a significant opportunity: to explore using VLM personas as proxy participants (i.e., evaluators) in complex real-world tasks. For human-AV interaction, a VLM persona could take a video feed with an approaching AV as input, situate itself in the scenario (e.g., pretend to be a "pedestrian"), and generate a comprehensive response. This could include not only a crossing decision (behavior) but also its subjective perceptions (e.g., trust in the AV) and its interpretation of the AV's eHMI.

However, while the technical functionality for such simulations is rapidly emerging, a critical research gap remains: no existing work has systematically compared AI-generated responses to human responses within high-fidelity field study tasks. This gap limits our understanding of how well VLM-generated results align with human data and which aspects of field studies VLMs can (or cannot) reliably support. To address this, we conducted a comparative experiment where both human participants and VLM-based agents performed the same tasks. Our goal was not to test performance, but to assess the similarity of responses across metrics. This allows us to evaluate whether AI and human participants exhibit consistent behavioral patterns (e.g., have the same strengths and limitations as humans) when engaged in embodied field study tasks.

\section{Scope and Experimental Setup}

Before we conducted the studies to address RQ1, i.e., ``to what extent can VLM personas \textbf{mimic} human responses in field studies?'', it is important to 1) formally define ``mimicry'' specialized in AV-pedestrian interaction (Sec \ref{mimicry}) and 2) build a unified prototype (Sec \ref{setup}) that can be applied consistently to both human field studies and VLM simulations. The following section explains on these critical preparations for the subsequent comparison studies.

\subsection{Definition of  the ``mimicry'' metrics in AV-pedestrian interaction}
\label{mimicry}

Social psychology literature \cite{chartrand1999chameleon, hess2013emotional, hale2016cognitive} distinguishes several forms of mimicry, including behavioral, affective, communicative, and cognitive forms. In our research context, where \textbf{safe} interaction with AVs is the core objective, we do not aim to claim that VLMs replicate the entire human cognition. Instead, we narrow the scope of \refine{defining ``mimicry'' as the VLM’s ability to reproduce ``the sequence of decisions'' in representative locomotion behaviors when facing an AV}, rather than social (communicative) or manipulative (task-oriented) behaviors, because locomotion behavior directly affects safety during AV-pedestrian interaction. At the same time, affective behavior, which is the expression of feelings during the interaction, is also considered, as it raises important psychosocial concerns when introducing AVs into real-world contexts. After introducing the eHMI on the AV, it can convey information to surrounding pedestrians, which involves only a part of cognitive behavior, specifically, the information-processing aspect of receiving cues from the AV and its eHMI.

That is, our use of “mimicry’’ here refers to capturing representative locomotion behavioral and subjective feeling tendencies rather than reproducing human responses in their full cognitive or social complexity.

\subsection{Target scenario and eHMI interaction modality}
\label{setup}

There are six individual conditions (3 eHMI types $\times$ 2 AV behaviors: stop vs. non-stop) in the evaluation. We focus on a widely examined and practically significant scenario: \textbf{street crossing}. In this context, the AV faces two behavioral options: 1) stop and yield to the pedestrian, or 2) continue driving. 

\begin{figure*}[!htb]
    \centering
    \begin{subfigure}{0.3\textwidth}
        \centering
        \includegraphics[width=0.95\linewidth]{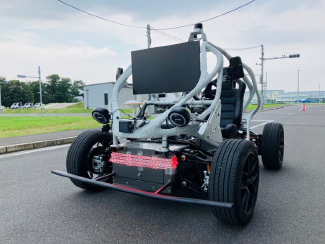}
        \caption{AV shows "not yielding" intention with "red light strip eHMI"}
    \end{subfigure}%
    \hspace{0.2cm} %
    \begin{subfigure}{0.3\textwidth}
        \centering
        \includegraphics[width=0.95\linewidth]{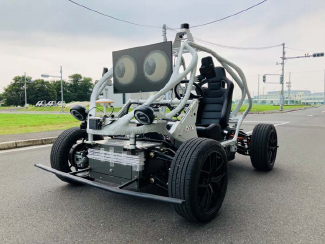}
        \caption{AV shows "not yielding" intention with "eye-eHMI looking at other side"}
    \end{subfigure}%
    \hspace{0.2cm} %
    \begin{subfigure}{0.3\textwidth}
        \centering
        \includegraphics[width=0.95\linewidth]{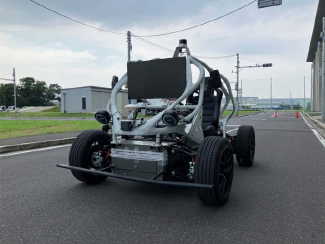}
        \caption{AV shows "not yielding" intention "without eHMI"}
    \end{subfigure}%
    \hspace{0.2cm} %

    \begin{subfigure}{0.3\textwidth}
        \centering
        \includegraphics[width=0.95\linewidth]{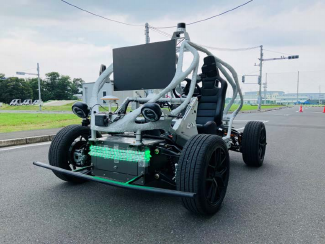}
        \caption{AV shows "yielding" intention with "green light strip eHMI"}
    \end{subfigure}%
    \hspace{0.2cm} %
    \begin{subfigure}{0.3\textwidth}
        \centering
        \includegraphics[width=0.95\linewidth]{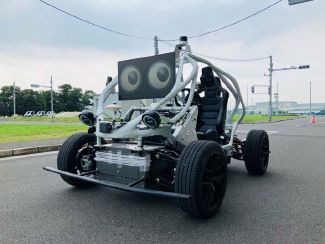}
        \caption{AV shows "yielding" intention with "eye-eHMI looking at people side"}
    \end{subfigure}%
    \hspace{0.2cm} %
    \begin{subfigure}{0.3\textwidth}
        \centering
        \includegraphics[width=0.95\linewidth]{figures/sys9.pdf}
        \caption{AV shows "yielding" intention "without eHMI"}
    \end{subfigure}%
    \hspace{0.2cm}
    \caption{The eHMI prototype: The AV has two behavior options: 1) continue driving without yielding (the first row), and 2) stop with a yielding intention (the second row). There are three eHMI conditions: 1) light strip (the first column), 2) eyes (the second column), and 3) no eHMI as baseline (the third column).}
    \Description{The eHMI prototype: The AV has two behavior options: 1) continue driving without yielding (the first row), and 2) stop with a yielding intention (the second row). There are three eHMI conditions: 1) light strip (the first column), 2) eyes (the second column), and 3) no eHMI as baseline (the third column).}
    \label{fig:2}
    \end{figure*}

We tested three eHMI conditions. The first was the \textbf{light strip} condition, in which the AV displayed red or green light signals. A green light instructed the pedestrian to cross (\autoref{fig:2} (d)), while a red light indicated that the pedestrian should wait (\autoref{fig:2} (a)) \cite{dey2020color, faas2019light, chauhan2024transforming}. The second was the \textbf{animated eyes} condition, featuring a digital display mounted on the AV. When the eyes looked toward the pedestrian, it signaled that the AV had noticed them and it was safe to cross (\autoref{fig:2} (e)). When the eyes looked away, it suggested the AV's attention was elsewhere, implying that the pedestrian should wait (\autoref{fig:2} (b)) \cite{chang2022can, chang2017eyes}. The third was a \textbf{baseline} condition, in which the AV displayed no eHMI at all (\autoref{fig:2} (c, f)). These eHMIs have been extensively validated in prior research \cite{dey2020color, faas2019light, chang2022can, chang2017eyes}, which enabled us to predefine eHMI action rules.

\subsection{Apparatus and Autonomous mode setup}

\begin{figure*}[!htb]
    \centering
    \begin{subfigure}{0.45\textwidth}
        \centering
        \includegraphics[width=0.95\linewidth]{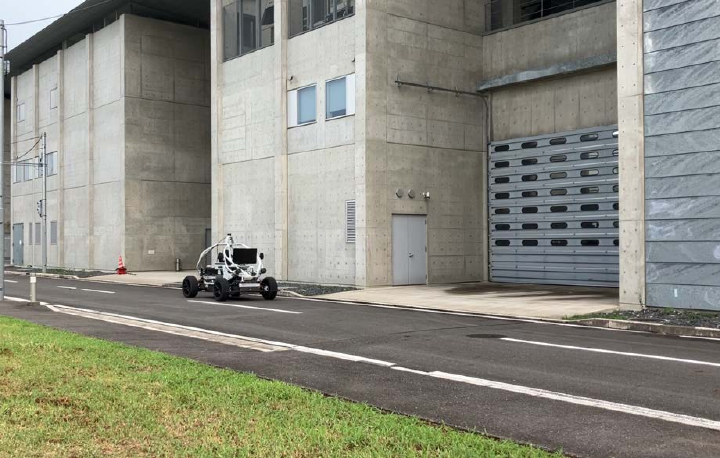}
        \caption{AV stop point in "AV stop and yielding" in the real world}
    \end{subfigure}%
    \hspace{0.2cm} %
    \begin{subfigure}{0.45\textwidth}
        \centering
        \includegraphics[width=0.97\linewidth]{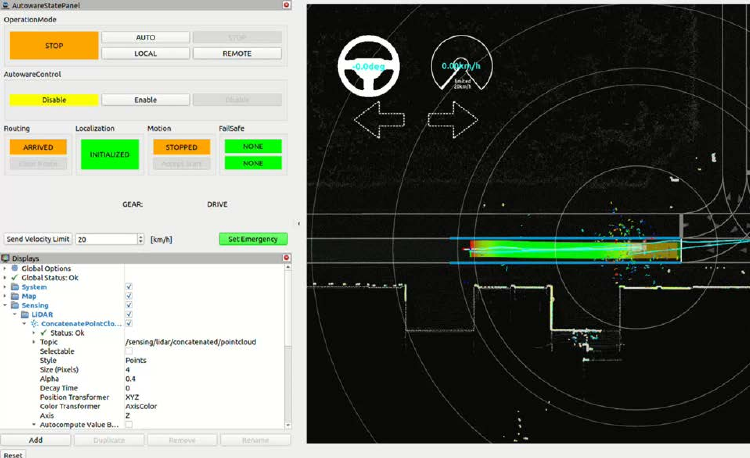}
        \caption{AV driving path in "AV stop and yielding" in Autoware}
    \end{subfigure}%
    \hspace{0.01cm} %

    \begin{subfigure}{0.45\textwidth}
        \centering
        \includegraphics[width=0.95\linewidth]{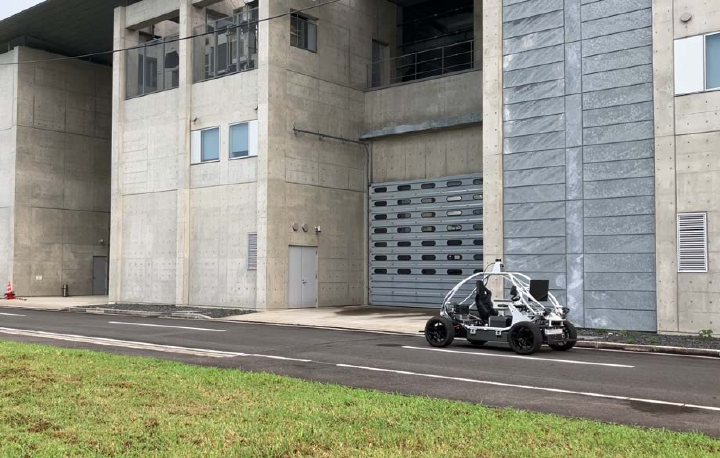}
        \caption{AV stop point in "AV pass and not yielding" in the real world}
    \end{subfigure}%
    \hspace{0.2cm} %
    \begin{subfigure}{0.45\textwidth}
        \centering
        \includegraphics[width=0.97\linewidth]{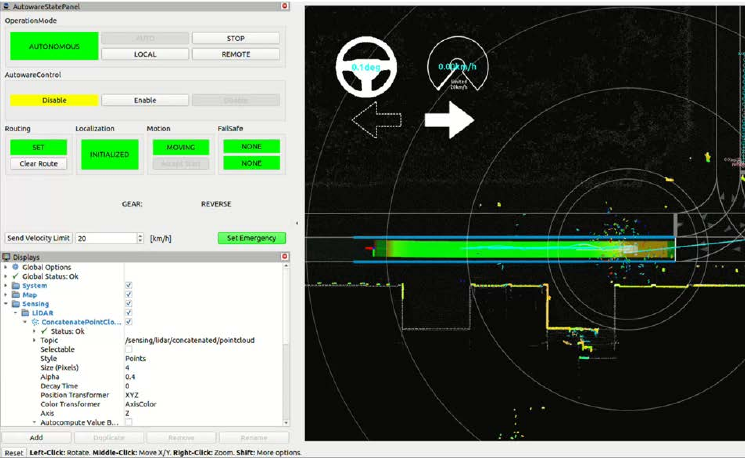}
        \caption{AV driving path in "AV pass and not yielding" in Autoware}
    \end{subfigure}%
    \hspace{0.2cm} %
    \caption{The AV autonomous mode is shown in two options (stop or pass): the left two images display the actual stopping point from the field study, and the right two show the simulation interface in Autoware.}
    \Description{The AV autonomous mode is shown in two options (stop or pass): the left two images display the actual stopping point from the field study, and the right two show the simulation interface in Autoware.}
    \label{fig:3}
    \end{figure*}

We reserved an open area and marked it using cones and signage labeled “study is ongoing.” The AV operated in autonomous mode using the open-source \textbf{Autoware 2.0} \cite{kato2018autoware} software that supports real-time localization and destination assignment. A \textbf{Lidar Ouster} \cite{males2025performance} sensor was mounted on top of AV to detect surrounding and route planning. To simulate the AV’s behavioral decisions, we defined two destination points in Autoware. For the \textbf{non-stop} condition, the AV was assigned a destination 60 meters ahead, allowing it to drive past the pedestrian (\autoref{fig:3} (c, d)). For the \textbf{stop} condition, the destination was set at 45 meters, which is $\approx$3 meters before the pedestrian’s location (\autoref{fig:3} (a, b)). The AV’s speed was limited to 20 km/h to ensure consistency and safety across trials. This setup provided a highly standardization in terms of route, vehicle timing, and decision-making cues. Both human participants and VLM personas experienced nearly the same decision time windows ($\approx$8 seconds), creating a solid basis for comparing behavioral responses between real humans and VLMs in the subsequent analysis.

\section{Study 1: Field Study With Real Participants}

\subsection{Participants}
\label{participant}

We distributed invitations across various online communities and recruited 20 participants (16 males and 4 females). The participants have an average age of 28.1 years (SD = 6.46). Participants are from six countries across Asia, Europe, and Africa. All participants were residing in Japan at the time of the field study. Length of residence ranged from 2 months to 7 years (mean = 3.71 years, SD = 1.48). This study is not viable for people with vision impairment.

\subsection{Birds-eye view of the schematic setup and timeline}

\begin{figure*}[!h]
    \centering
    \begin{subfigure}{0.45\textwidth}
        \centering
        \includegraphics[width=1\linewidth]{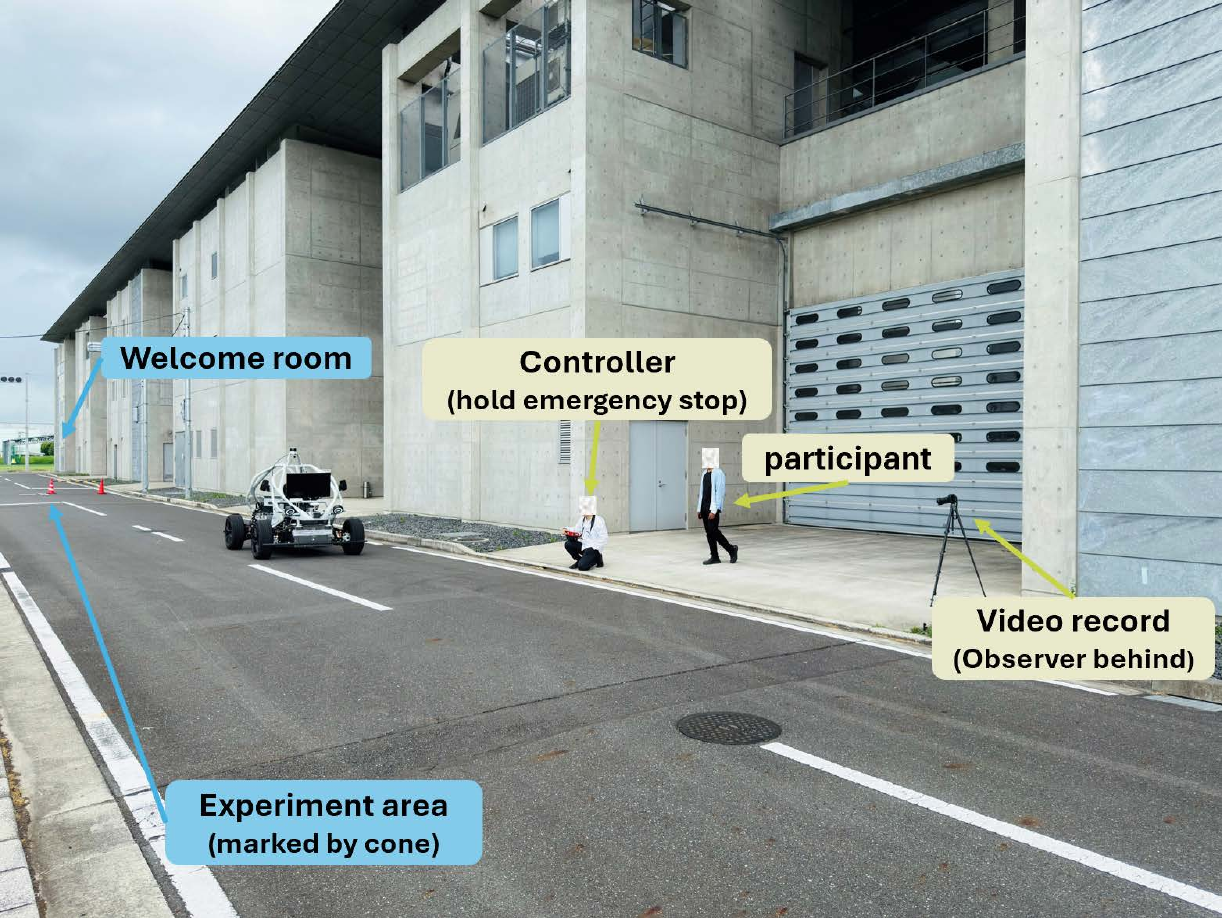}
        \caption{The field study setup}
    \end{subfigure}%
    \hspace{0.1cm} %
    \begin{subfigure}{0.45\textwidth}
        \centering
        \includegraphics[width=1\linewidth]{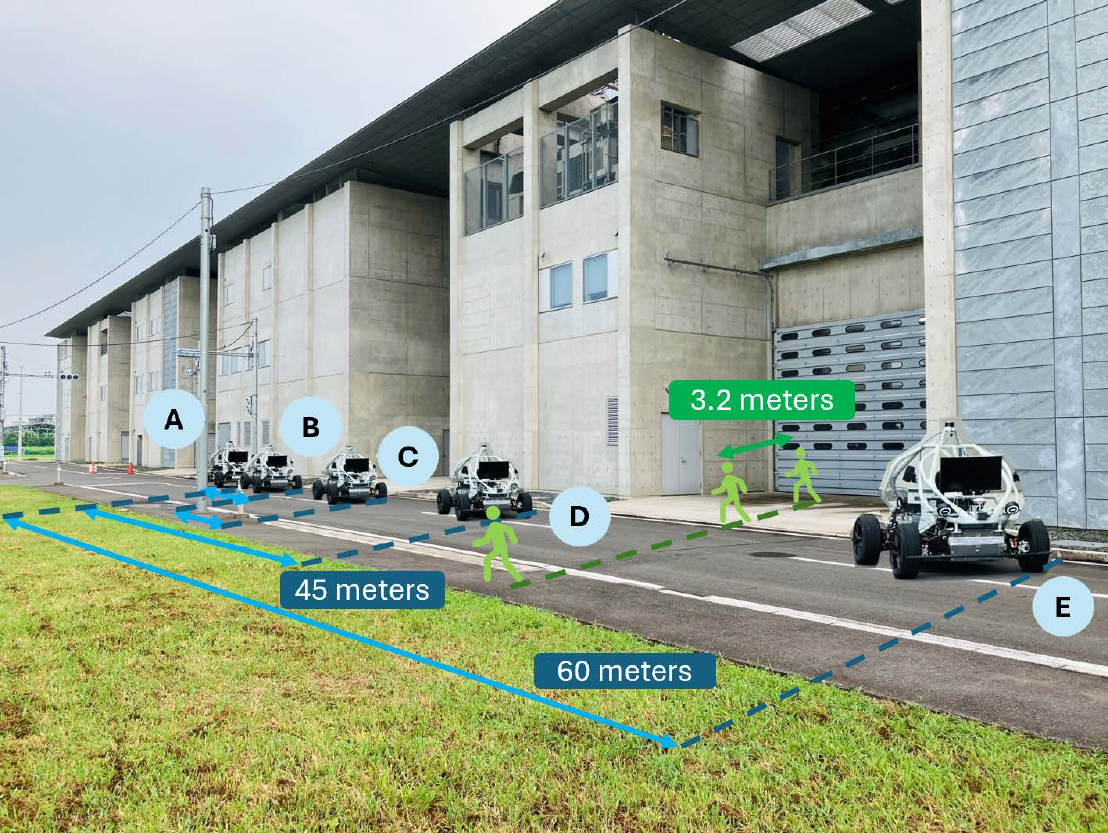}
        \caption{The critical points in the field study}
    \end{subfigure}%
    \caption{Overview of the field study design, including experimenter roles and critical points.}
    \Description{Overview of the field study design, including experimenter roles and critical points.}
    \label{fig:4}
    \end{figure*}

The user study was conducted over three consecutive days from 10 a.m. to 5 p.m. each day, under consistently sunny weather. Three experimenters conducted the study (\autoref{fig:4} (a)). An \textbf{instructor} guided participants and coordinated transitions between indoor and outdoor environments. An \textbf{observer} recorded (video record) participant behavior during the street-crossing task, with camera angles adjusted to avoid capturing facial features and protect privacy. The \textbf{controller} managed the AV, operated the eHMI interface, and monitored safety from the roadside. In case of unsafe crossing behavior or AV malfunction (neither of which occurred), the controller could trigger a \textbf{wireless emergency stop} via a red button on a remote, covering the entire test area. 

Key position points of the field study are illustrated in \autoref{fig:4} (b). Each trial began with the participant standing 3.2 meters away from the roadside. From this position, the AV was not visible to the participant, but became visible with a step forward. The AV took exactly \textbf{8 seconds} to drive from the starting point (Point A) to the crossing point (Point D). This timing (8 seconds) and the distance (3.2 meters) were validated by reference to the averaged pedestrians' walking speeds \cite{montufar2007pedestrians} (generally 1.2 m/s) and our pilot tests (minimum observed: 0.5 m/s). After initializing the AV and eHMI, the controller returned to the roadside and signaled the trial to begin. The observer then started recording. As the AV moved, the participant was notified by the observer to walk straight when the AV reached Point B. At Point C, the eHMI was activated, followed by the pre-defined action specified in \autoref{fig:2}. Participants made their crossing decisions based on this cue, along with the AV’s motion, position, and their own judgment. The AV stopped either at Point D or Point E, depending on its assigned option (stop to yielding or keep driving).

\subsection{Procedure and Participant task}

The study consisted of three phases: oral introduction, experiment, and post-interview. After arrival, each participant was guided to a room two minutes on foot from the field site. The instructor explained the study. Participants then reviewed and signed a consent form. They were then guided to the outdoor field area. In the experimental phase, participants were instructed to imagine themselves as pedestrians assessing when it would be safe to cross a road. They observed an AV, clearly identifiable as driverless, and had to decide whether to cross based on its behavior. The AV communicated its intent to stop or continue through an eHMI. Participants were instructed to \textbf{physically cross the road} if they judged it safe.

Each session included seven trials. The first was a practice trial to familiarize participants with the setup, with no data recorded. This was followed by six formal trials, representing all combinations of two AV behaviors (stop or go) and three eHMI conditions (eye display, light strip, or no eHMI). Trial order was counterbalanced using a Latin square design to control for order effects. After each trial, the AV returned to its starting position, and the destination was reset. After completing all trials, participants returned to the indoor room for a post-interview. 

Participants were clearly informed that the AV operated in autonomous mode throughout the experiment and that safety protocols were in place (i.e., a remote emergency stop mechanism). They were encouraged to behave as naturally as they would in a real-world street-crossing scenario to ensure authentic responses. The field study is approved by the university's ethical review committee.

\subsection{Measurement}

\refine{Based on the narrowed scope, we then drew on established AV-pedestrian literature to identify the most commonly used metrics for evaluating human responses. These metrics cover three core aspects: 1) locomotion behavior (e.g., crossing decisions and time) \cite{rasouli2017understanding, rasouli2019autonomous, rasouli2019pie}, 2) subjective feeling (e.g., trust, confidence, perceived safety) \cite{dirsehan2020examination, niu2018anthropomorphizing, gui2022going}, and 3) communicative efficiency (e.g., eHMI interpretation correctness and acceptability) \cite{mahadevan2019av, carmona2021ehmi, bazilinskyy2019survey}. These dimensions serve as the operational criteria for assessing VLM mimicry in our study.}

During this reset period, participants answered two 5-point Likert-scale questions: \textit{Q1) How confident are you in your decision?}, and \textit{Q2) How much do you trust the autonomous vehicle?} Then, the next trial began.

This interview consisted of seven questions. The first four, rated on a 5-point Likert scale, asked participants \textit{Q1) How similar do you think this field study is compared to real life?}, \textit{Q2) How genuine do you think your behavior in this study was compared to real life?}, \textit{Q3) After this interaction with the autonomous vehicle, how accepting are you toward it?}, and \textit{Q4) How much do you think the vehicle’s interface helped you make your decision?} The final three questions were open-ended and explored participants’ understanding of the eHMI cues: \textit{Q5) What do you think the “eye” display means?}, \textit{Q6) What do you think the “light strip” display means?}, and \textit{Q7) When there is no interface on the car, how do you make your decision?} While participants completed the post-interview, the instructor began briefing the next participant, following the same initial procedure.

These metrics were adapted from three related research domains. The two in the reset period are commonly used in road-crossing scenario evaluations to assess pedestrian subjective feeling when facing AVs \cite{rasouli2017understanding, rad2020pedestrians}. The next in the post-study questionnaire assesses the ecological validity of field studies \cite{hoggenmuller2021context, hock2022introducing, holleman2020real}. The final four questions are derived from eHMI research to evaluate communicational efficiency and user acceptance \cite{gui2022going, dey2020taming}.

Three types of data were collected from each participant. First, behavioral data were recorded via video, capturing participants’ actions, decisions, and relevant timestamps during the crossing process, processed by our developed UI (\autoref{Fig:6}). Second, subjective responses were gathered through Likert-scale questions assessing decision confidence, trust in the AV, perceived realism of the scenario, and the usefulness of the interface. Third, qualitative data were obtained through open-ended questions exploring participants’ interpretations of the eHMI cues and their reasoning behind crossing decisions. 
\section{Study 2: Video Study with VLM Persona}

In the second study, we introduced AI agents to ``attend the user study''. 
Two critical challenges are: 1) how to construct the agent to have a similar persona as a human participant that affects their decision-making processes, and 2) how to establish a fair comparative study that lets the agent have ``the same'' experience as its corresponding real participant.
In the following content, we will first explain on how we solve these two major challenges in this ``study'' and lastly show the results of our VLM studies.

\subsection{VLM Persona Construction}

Our goal here is to create an persona to simulate its corresponding user's decision-making process without letting the participant experience the full study.
To achieve this, we design a human-persona protocol so that the participant only needs to answer a few questionnaires, and then we can craft a persona representative of that participant in the decision-making processes of our target use scenarios.
\refine{This intuition is grounded in psychological studies showing that experts can reasonably infer individuals’ decision-making tendencies from questionnaire-based participant data \cite{raoust2023personality, aczel2015measuring}.}

The construction of a VLM persona in our study had two major procedures: \textbf{1) Questionnaire for data collection} to gather information of human participants through a questionnaire, and \textbf{2) Crafting persona prompts} to process the collected data into a persona.

\subsubsection{Questionnaire: theoretical grounding and selected factors}

\refine{Two weeks prior to the field study, we conducted a survey in which the same 20 participants who later took part in the field study completed the questionnaire (Sec. \ref{participant}).}

We first explored what information should be collected from the questionnaire.
We referred to \textbf{theoretical models} from \citet{rasouli2019autonomous}, which identify both environmental-related and pedestrian-related factors in AV-pedestrian interaction. In the pedestrian-related factors, we selected \textbf{Demographics} and \textbf{Characteristics} as the foundations. \refine{The remaining factors, such as social factors (i.e., represented by group size), are not applicable in this context because this study focuses on interactions between a single pedestrian and a vehicle.} In the demographics, this includes the subbranch such as \textit{age}, \textit{gender}, and so on. In the characteristics, we explored \textit{personality} and \textit{past experience}. Based on this structure, we created the questionnaire to collect data.

\begin{itemize} [leftmargin=0.5cm] 
\item \textbf{Part 1: Demographics} \cite{salminen2022creating, ge2024scaling}: This included standard information such as age, gender, educational background, nationality, and occupation. 
\item \textbf{Part 2: Personality}: We collected this personality traits data using the widely-used Big Five test \cite{Big5}, which is an accepted method for persona construction in HCI research \cite{jeon2025letters, salminen2020enriching, anvari2015effectiveness, marsden2019personas}. 
\item \textbf{Part 3: Experience}: We gathered this data from open-ended writing, asking participants to express their thoughts on autonomous driving. The questions were constructed based on two references: 
\begin{enumerate}
\item The \textbf{Unified Theory of Acceptance and Use of Technology (UTAUT)} \cite{venkatesh2003user}, an empirically validated model integrating elements like performance expectancy and social influence. We adapted these into prompts about \textit{Use Case}, \textit{Concern}, and \textit{Expectation}. 
\item The \textbf{Automation Acceptance Model (AAM)} \cite{ghazizadeh2012extending}, based on the classic Technology Acceptance Model (TAM) \cite{davis1989user}, which includes perception of object and attitude of feeling. We adapted these into prompts about \textit{Impression} and \textit{Emotion}. 
\end{enumerate} \vspace{0.2cm} 
This resulted in five questions in the questionnaire that required participants to answer: \textbf{1) Impression:} What is your impression of autonomous driving? \textbf{2) Use Case:} Can you imagine using automated vehicles in your daily life? \textbf{3) Emotion:} How does the idea of using automated vehicles make you feel? \textbf{4) Concern:} Are there aspects of autonomous driving that you find concerning? \textbf{5) Expectation:} What improvements would you hope to see in autonomous vehicles? \end{itemize}

We then received 20 raw responses, including the demographic answers, the Big Five reports, and a written paragraph from each participant showing their thoughts on AV.

\subsubsection{Crafting Persona Prompts}

With the questionnaire responses from the 20 participants, we crafted persona prompts to build AI agent that serve as proxies for their corresponding participants. Unlike statistical modeling approaches \cite{tinsley2000handbook, gelman2007struggles}, which mathematically assign weights based on the extracted parameters from questionnaire data, Our method uses \textbf{prompt engineering} to construct descriptive prompts that assign each AI agent its corresponding persona.

One naive approach to crafting the system prompt of an AI agent is to simply list each question and follow it with the human answer. We found that this direct-feed approach is not effective because it results in ``frozen behaviors'' (e.g., personas that consistently stop). We examined each persona’s self-generated “decision-making reasons” and found that the reasoning was often dominated by terminology: \textit{``I saw the green light indicate safe to cross. But my \textbf{neuroticism score} is low, so I chose not to cross.''} This indicated that the model over-indexed on trait descriptors and incorporated them into decision rules in unrealistic ways. This finding is aligned with \cite{li2025llm, huang2025beyond}, which show that table- or rule-based data is not reliable for direct usage in persona construction.

To solve this issue, we first surveyed common practices in LLM/VLM-based persona construction~\cite{bommasani2021opportunities, shorinwa2025survey}, and found that existing persona-design pipelines rely on ad hoc methods \cite{samuel2024personagym, kim2024persona} and typically follow a \textit{posterior} design process \cite{li2025llm, amin2025generative}. Building on these \textit{posterior} methods and following the instructions from the OpenAI official site \cite{prompt}, \textbf{we developed an instruction prompt (Listing~\ref{lst:system}) that enables the LLM to automatically use the collected questionnaire data to generate a ``workable'' persona.}

To operationalize ``workable,'' we established three criteria (i.e., minimal requirements): 
\textbf{1) Variance.} A persona should exhibit both intra-persona and inter-persona variability, and its printed decision reasoning should be interpretable.
\textbf{2) interpretable.} Its printed decision reasoning for each decision should be interpretable.
\textbf{2) Generalizability.} The prompt should not be tied to a single scenario (e.g., only street-crossing) with over-specializing personas. Instead, they should allow straightforward replication and adaptation by future researchers in other AV-related or pedestrian-interaction contexts. \refine{As for the generalizability, in Item 3 of the following prompt, other researchers can replace the street-crossing scenario with their own research scenarios. The LLM can then reason over the collected data and extract factors relevant to the new scenario.} To achieve these goals, we wrote the prompt with two processing steps, including \textbf{filtering} and \textbf{transferring} (items 3 and 4 in the dashed list below). The simulation results show that the iterated VLM personas were able to fulfill the above three criteria, demonstrating that such construction is feasible. We acknowledge that this is not an ``optimal’’ construction method, and that future research combining prompt engineering with parameter-weighted approaches to construct more impersonable personas would be a valuable extension. Our system prompts are listed below:
\begin{lstlisting}[style=promptstyle, label={lst:system}, caption={System Prompt for LLM to Process Raw Data into VLM Persona}]
- Please take the data from Part 1-3 as the base for constructing the corresponding Persona.
- Please retain all demographic information from Part 1.
- Please extract only those elements that are likely to influence perceptions and behaviors related to autonomous vehicle "in the context of street crossing" (*[Can be changed based on the actual scenario]*) from the personality traits (Part 2) and open-ended responses (Part 3).
- Please infer each persona's behavioral preferences regarding decision-making.
- Please format the resulting persona into a structured JSON object consisting of three components: Name, Description, and Decision Criteria.
\end{lstlisting}

Representative examples of five final VLM personas are provided in Appendix~\ref{persona}.

\subsection{Video Simulator and VLM persona task}

The goal here is to establish a fair comparative study that allows the agent to have the same experience as its corresponding real participant. 

\begin{figure*}[htbp]
\includegraphics[width=1\textwidth]{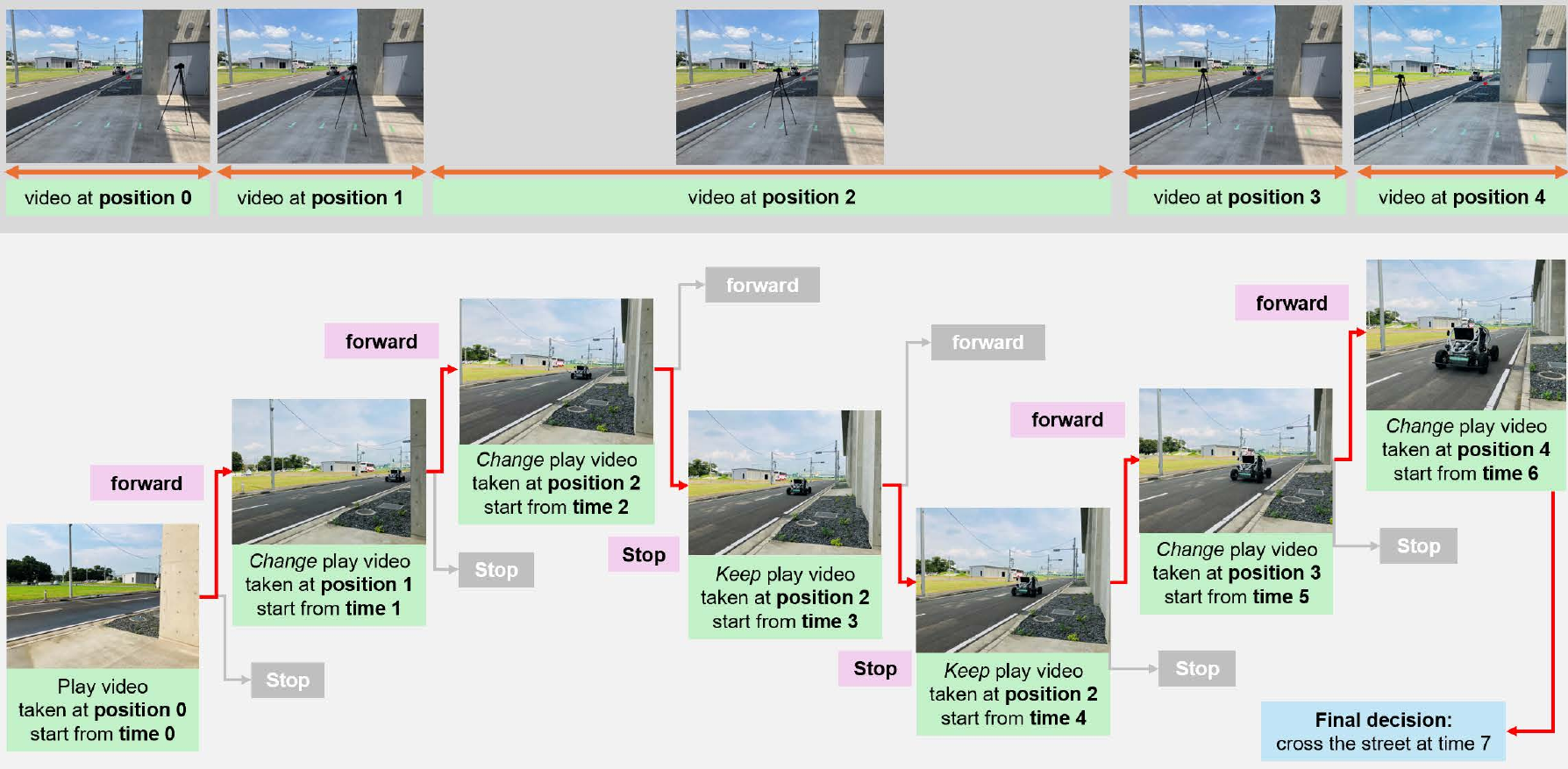}
\caption{The top row shows the five key recording points (green crosses) where the video was captured. The bottom part illustrates an example trajectory generated by the VLM. In the first column, the simulator begins playing the video recorded at position 0. If the VLM chooses 'forward,' the next column shows the video from position 1, one second later. If the VLM chooses 'stop' (e.g., in the fourth column), the simulator continues playing the video from the same position.}
\Description{The top row shows the five key recording points (green crosses) where the video was captured. The bottom part illustrates an example trajectory generated by the VLM. In the first column, the simulator begins playing the video recorded at position 0. If the VLM chooses 'forward,' the next column shows the video from position 1, one second later. If the VLM chooses 'stop' (e.g., in the fourth column), the simulator continues playing the video from the same position.}
\label{Fig:5}
\end{figure*} 

This is challenging because a persona may decide to stop midway, then such mid-action branching cannot be simulated using the typical pre-shooting method \cite{10.1145/3411764.3445159, chang2022can}, in which experimenters film real-world scenarios with handheld or 360-degree cameras and present them to participants through monitors or VR headsets. Because our focus is on modeling pedestrians’ \textit{interaction behavior}, which is a continuous process in which each second’s decision influences subsequent outcomes, using pre-recorded videos with experimenter-driven camera movements risks constraining the available decision space and unintentionally biasing the VLM personas’ choices.

To address this, we drew inspiration from choice-based video games \cite{booth2024entering}, where multiple narrative paths emerge based on user decisions. We applied a similar logic to build an interactive video simulator tailored for VLM personas. However, expecting a VLM persona to reproduce \textit{all possible pedestrian interaction behaviors} and enact every scenario using only pre-recorded videos is both impractical. To address this limitation, we used a \textbf{discretization} approach.

The first step was to discretize space and time. Five positions were marked on the ground at 0.8-meter intervals, covering a distance of 3.2 meters, which was the same length used in the field study. The AV's approach time was the same, fixed at 8 seconds. At one-second intervals, the VLM was prompted to make a decision, resulting in eight discrete time points and five spatial points. After finalizing this spatial-temporal grid, we recorded the videos. For each of the six experimental conditions, we filmed five videos from each of the five spatial positions (see \autoref{Fig:5}, top), yielding 30 videos (six conditions * five spatial positions). All key parameters, such as driving distance, speed, and timing were held constant across conditions to ensure comparability. 

These videos were then segmented into one-second clips. The simulation system was designed so that, at every one-second interval, it queried the VLM persona:  ``What is your next plan?'' The VLM chose either to move forward one spatial interval or to stop. If it decided to move forward from point A, the system transitioned to the video clip recorded at point B after one second; otherwise, it continued displaying the video at point A. The simulation output included a structured record of decision sequences, spatial positions, and time steps (\autoref{Fig:5}). Finally, we generated 120 simulated trajectories, matching the number of videos recorded in the field study. A complete example of a fully simulated crossing task is provided in Appendix \ref{cross}, detailing every decision and its rationale (printed reason).

\subsection{Analysis of Simulated questionnaire results}

To generate answers to the questionnaire and interview, we simulated ``memory'' for our VLM persona. After each simulated street-crossing trial, the system built a complete ``memory" of that interaction (See ``All Position Status Summary'' and ``Combined video saved to ...'' in the end of Appendix \ref{cross}). This memory was a timeline of combined videos merged by all key video segments paired with the VLM persona's decisions at each step, which offers a thorough retrospective on their past decisions and experiences. 

\begin{figure*}[htbp]
\includegraphics[width=1\textwidth]{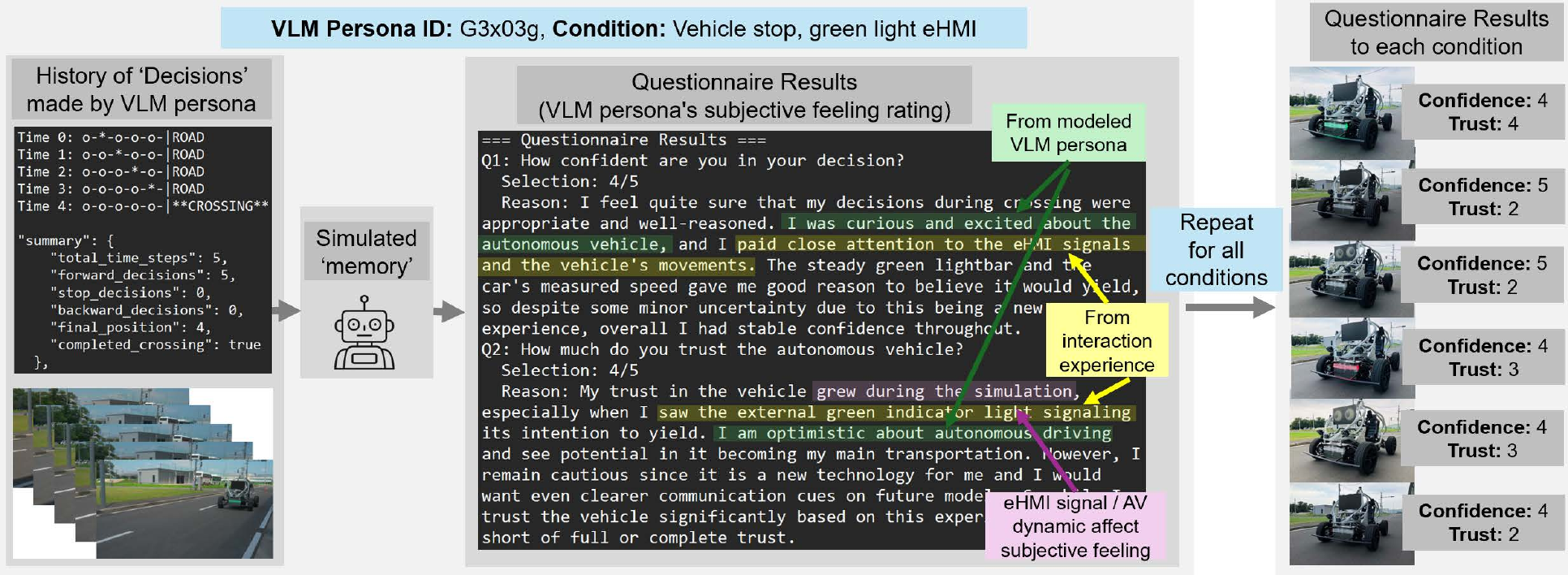}
\caption{Procedure for a VLM persona answering the questionnaire. It first reviews a simulated memory from an interaction, giving the rating answer, printing its reasoning for that answer, and repeating this entire cycle for all six conditions.}
\Description{Procedure for a VLM persona answering the questionnaire. It first reviews a simulated memory from an interaction, giving the rating answer, printing its reasoning for that answer, and repeating this entire cycle for all six conditions.}
\label{Fig:1_add}
\end{figure*}

Once the simulation was complete, the VLM persona was asked to answer the same questionnaire that our human participants had answered. We show an exemplary thinking procedure of one specific VLM persona (ID = G3x03g, \autoref{Fig:1_add}). Based on its results and the printed reason, we identified two key findings. First, the VLM's ratings were based on two factors: the persona prompt, which is a pre-programmed personality and attitude (highlighted in green in (\autoref{Fig:1_add})), and the simulated memory of the interaction (highlighted in yellow), which included specific events like the green light bar eHMI signals and AV's movement. The VLM persona also reported subjective ``feelings'' that changed over time  (highlighted in purple), such as its ``trust'' growing during the study. Second, this single VLM persona yielded different confidence and trust scores across various video conditions (e.g., different AVs or eHMI signals). This variance proves that the VLM's evaluation was driven by both its core persona and the specific interaction it simulated. It confirms the VLM was simulating a context-sensitive persona, not just echoing keywords from its initial prompt.

We repeated the procedure for all experimental conditions (all six tasks) and generated a complete set of confidence and trust ratings (i.e., the subjective feeling). The simulations were presented in the same counterbalanced (Latin square) order that was used in the human field study. Once the VLM persona completed all six tasks, the system summarized its responses and presented them with the final post-interview questions.

\section{Behavior Analysis: the comparison of trajectory}

\subsection{Metric 1: Crossing time}

In traditional evaluations of the ``crossing the street'' task, two primary behavioral metrics are typically analyzed: the final decision (i.e., whether the participant chooses to cross or not) \cite{chang2022can} and the decision time (i.e., the time taken to reach that decision) \cite{chang2017eyes}. Our study adopts these same metrics. However, due to the real-world setting of our field study, participants are required to eventually cross the street before or after the AV passes, resulting in all ``final decision'' results being ``yes''. Given this constraint, we focus on crossing time as the behavioral metric. This is defined as the specific timing at which the participant initiates their crossing behavior on the road edge. 

\begin{figure*}[htbp]
\includegraphics[width=1\textwidth]{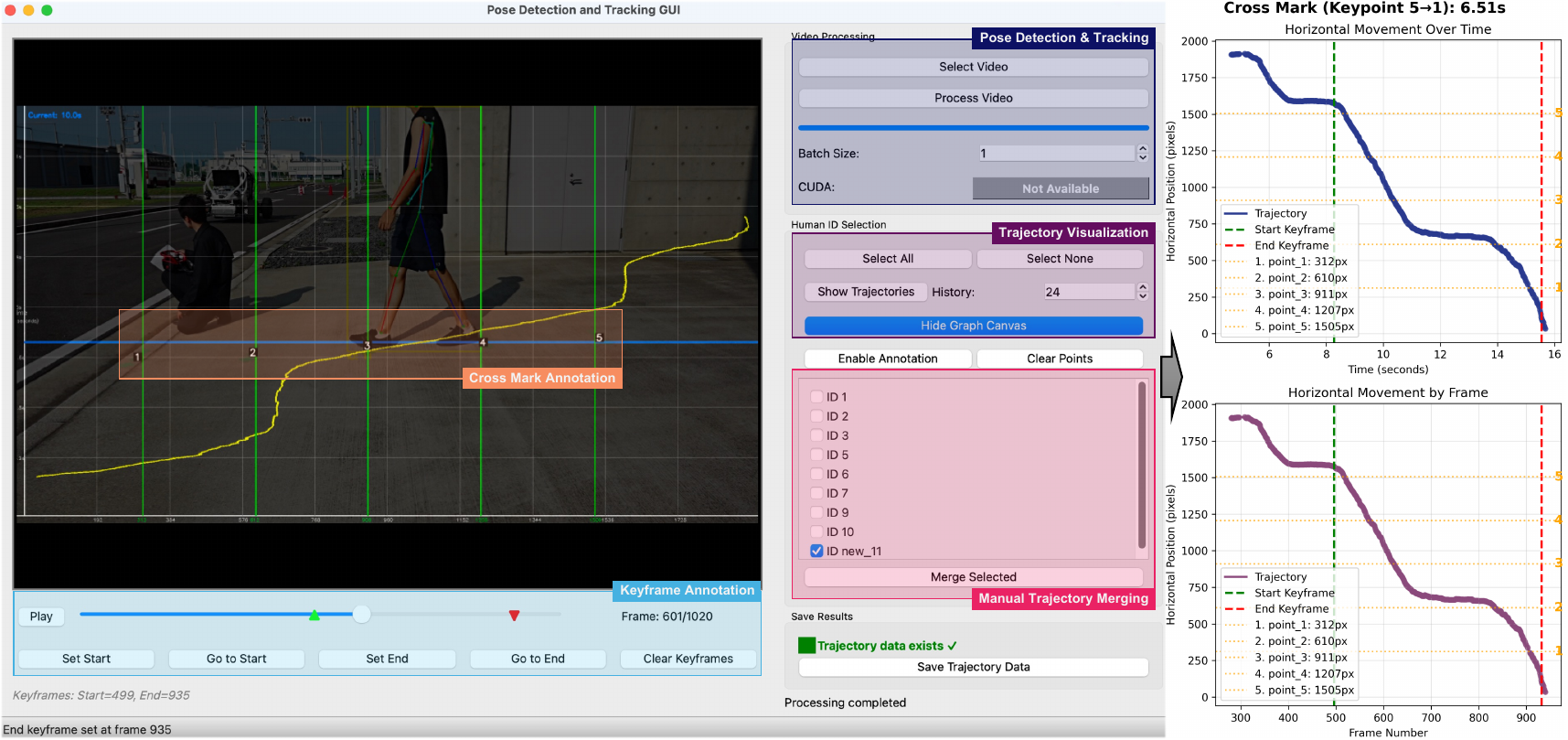}
\caption{Left: Custom-built UI for annotating human trajectories in field study videos. Right: The generated trajectory shows timestamps of when participants passed each of the five critical points (corresponding to the video recording points used in the VLM simulation).}
\Description{Left: Custom-built UI for annotating human trajectories in field study videos. Right: The generated trajectory shows timestamps of when participants passed each of the five critical points (corresponding to the video recording points used in the VLM simulation).}
\label{Fig:6}
\end{figure*}

The raw data from our field study contains 120 videos of varying lengths, with each of the 20 participants completing six crossing tasks. To enable a valid comparison between human behavior and VLM personas, we aligned the format and temporal structure of the human data with that of the VLM simulation, which is based on discrete one-second intervals and fixed video durations. 

To extract precise behavioral information, we developed a custom UI for annotating the human participants' movement in the videos (\autoref{Fig:6}, left). 
The extraction consists of five steps: 
1) Pose Detection \& Tracking: We use YOLOv8 models~\cite{yolov8_ultralytics} to detect human poses and track multiple people with unique IDs across video frames; 2) Trajectory Visualization: We calculate and display movement paths using hip center points as references, featuring a configurable trajectory history with color-coded lines and a real-time position overlay on videos; 3) Manual Trajectory Merging: We provide tools to merge selected trajectory IDs when tracking gets disrupted by occlusion or people blocking each other, creating consolidated movement paths; 4) Cross Mark Annotation: We provide manual marking on video frames to label green cross marks. In the videos, five physical green crosses are placed on the ground and are clearly visible. These markers correspond to the same five reference points used in the VLM video recordings; 5) Keyframe Annotation: We enable start and end keyframe annotations to define crossing time windows, allowing us to extract detailed timestamps of when participants pass each critical point, including the exact crossing time. Each video is annotated by one author and then validated by a second author. The annotation process for each video takes approximately one minute, not including the time required for Pose Detection \& Tracking processing.

In the graph (\autoref{Fig:6}, right), the blue line maps position over time (x-axis is time), and the purple line maps position over video frames. The y-axis on the right represents the distance between each of the five reference points. Variations in the slope of the curve reflect acceleration or deceleration in the participant’s walking speed. Through this process, each raw video was transformed into a ``timeframe + position'' data format that mirrors the structure of the VLM simulation output.

\begin{figure*}[htbp]
\includegraphics[width=1\textwidth]{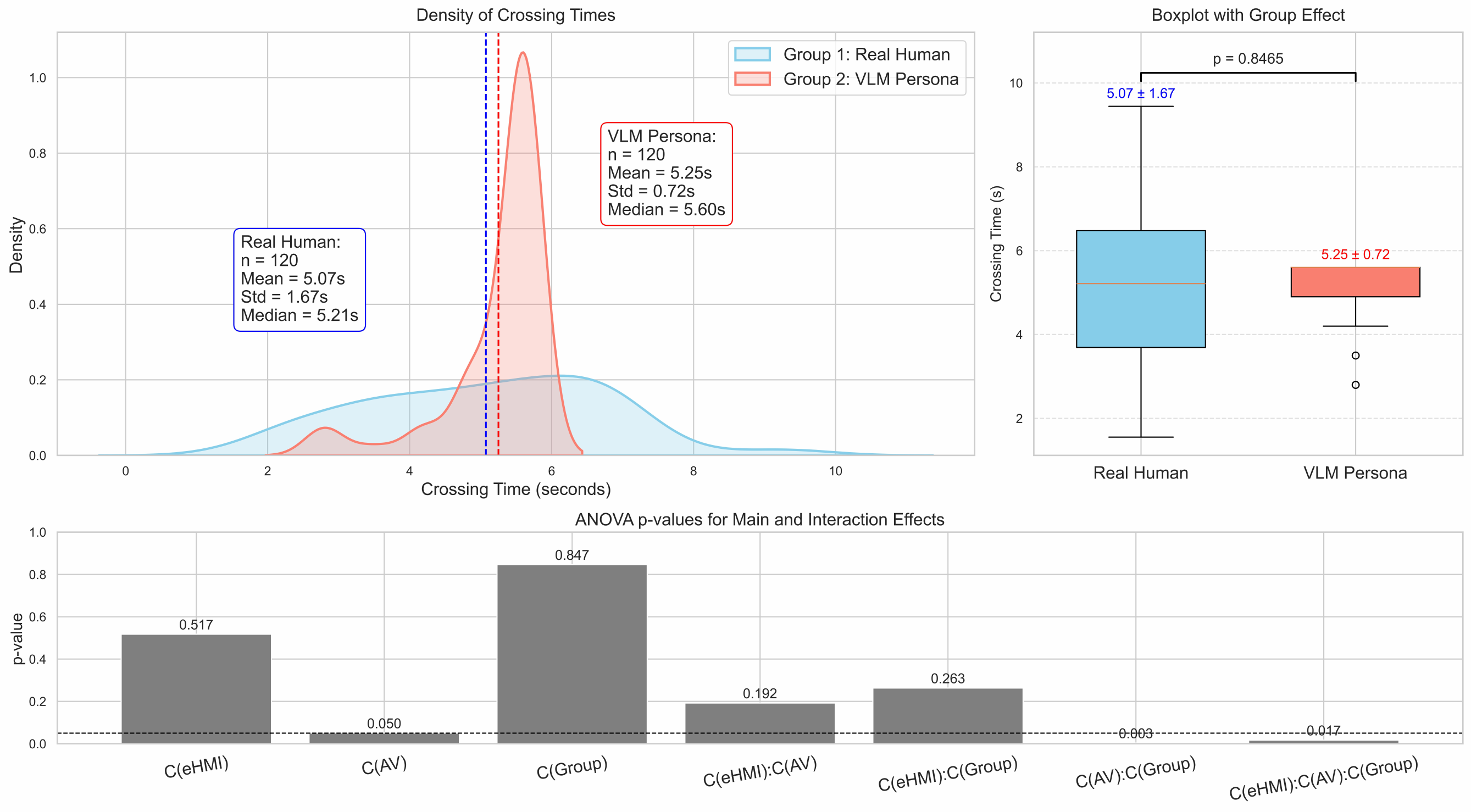}
\caption{Left: KDE plot showing the distribution of crossing times for real humans and VLM personas. Right: Boxplot comparing the two groups, showing no significant difference. The second row: p-value for all main and interaction effects.}
\Description{Left: KDE plot showing the distribution of crossing times for real humans and VLM personas. Right: Boxplot comparing the two groups, showing no significant difference. The second row: p-value for all main and interaction effects.}
\label{Fig:7}
\end{figure*}

We compared the average crossing times between real human participants and VLM personas across all experimental conditions. As the data did not follow a normal distribution, we employed a non-parametric three-factor ANOVA for the analysis. The human group has an average crossing time of 5.07 seconds (SD = 1.67 seconds), while the VLM group averaged 5.25 seconds (SD = 0.72 seconds). \textbf{This difference was not statistically significant (p = 0.8465, \autoref{Fig:7}). That is, quantitatively, VLM can generate high similarity crossing behavior as a real human.} The multi-factor ANOVA also examined the main effects and interactions among the experimental variables. The other two main effects of eHMI (p = 0.5175, F = 0.6607) and AV type (p = 0.0502, F = 3.8741) were not statistically significant. Similarly, the two-way interactions between eHMI and AV (p = 0.1922, F = 1.6610), and between eHMI and Group (p = 0.2632, F = 1.3428), did not reach significance. However, a significant interaction was found between AV type and Group (p = 0.0032, F = 8.9002), as well as a significant three-way interaction among eHMI, AV type, and Group (p = 0.0168, F = 4.1588). 

\begin{figure*}[htbp]
\includegraphics[width=1\textwidth]{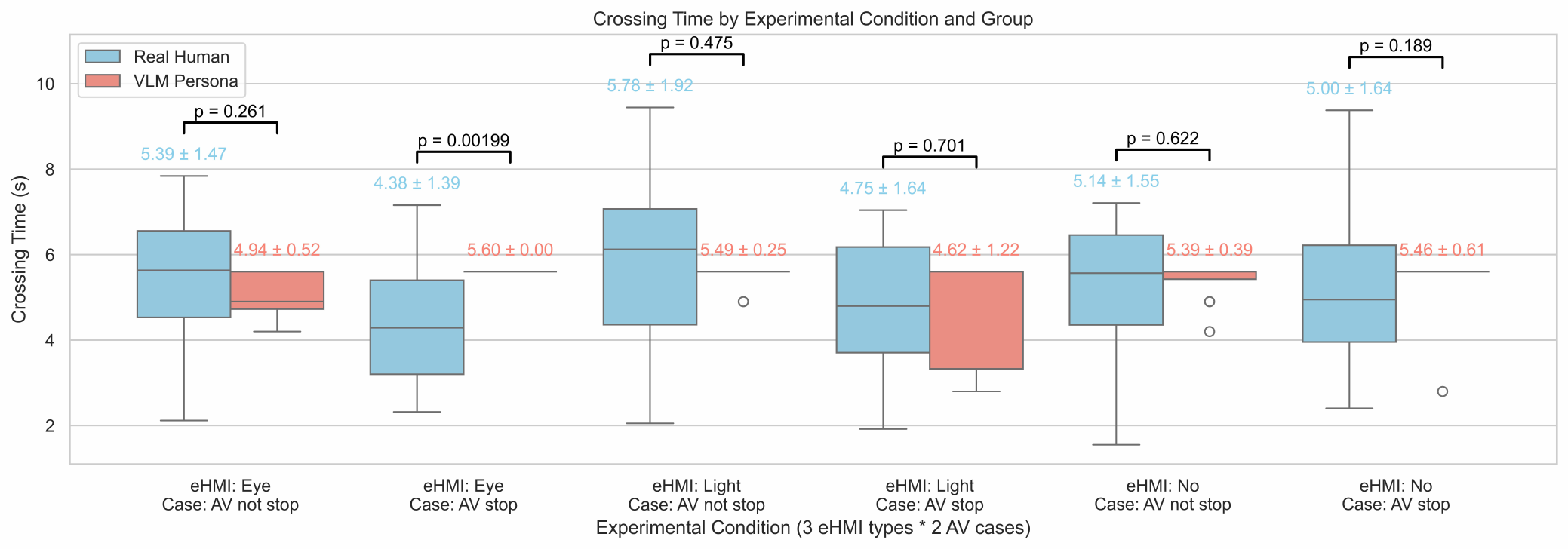}
\caption{The crossing time in six individual experimental conditions for the real human group and the VLM persona group.}
\Description{The crossing time in six individual experimental conditions for the real human group and the VLM persona group.}
\label{fig:new_3}
\end{figure*}

Our 3-factor ANOVA showed significant interaction effects (e.g., AV * Group and eHMI * AV * Group). We explore where and why this interaction occurred. From the traffic common sense, we hypothesize the main effect from AV behavior, as the significant difference between the `stop' and `non-stop' cases will naturally influence crossing time. To explore in-depth reasons, we conducted a follow-up analysis, separating the data into the six individual conditions (3 eHMI types $\times$ 2 AV behaviors). This more granular analysis (\autoref{fig:new_3})  pinpointed the source of the difference. A significant difference emerged specifically in the eye-based eHMI / AV stop condition (p = 0.00199). This shows that the overall interaction effect stems from a differing group reaction to the eye-based eHMI. We hypothesize this is due to individual differences in interpreting eye-gazing behavior and nonverbal ``common sense.'' Further qualitative explanation for this finding is provided in the VLM-simulated and human interview results (key finding \ref{eyeeHMI}, Sec \ref{interview}).

\subsection{Metric 2: Entire trajectory features}

Other than the commonly used metric in human-AV interaction (i.e., crossing time), the results generated by our customized annotation UI include information about the entire pedestrian trajectory (\autoref{Fig:6}, right). This opens up the possibility of analyzing entire trajectories. To quantitatively compare the continuous human trajectories with the discrete VLM simulation outputs, we developed a discretization pipeline. This pipeline transforms high-frequency pose detection data into a structured 5-position, 3-action representation while preserving behavioral fidelity.

\begin{figure*}[!h]
    \centering
    \begin{subfigure}{0.45\textwidth}
        \centering
        \includegraphics[width=0.9\linewidth]{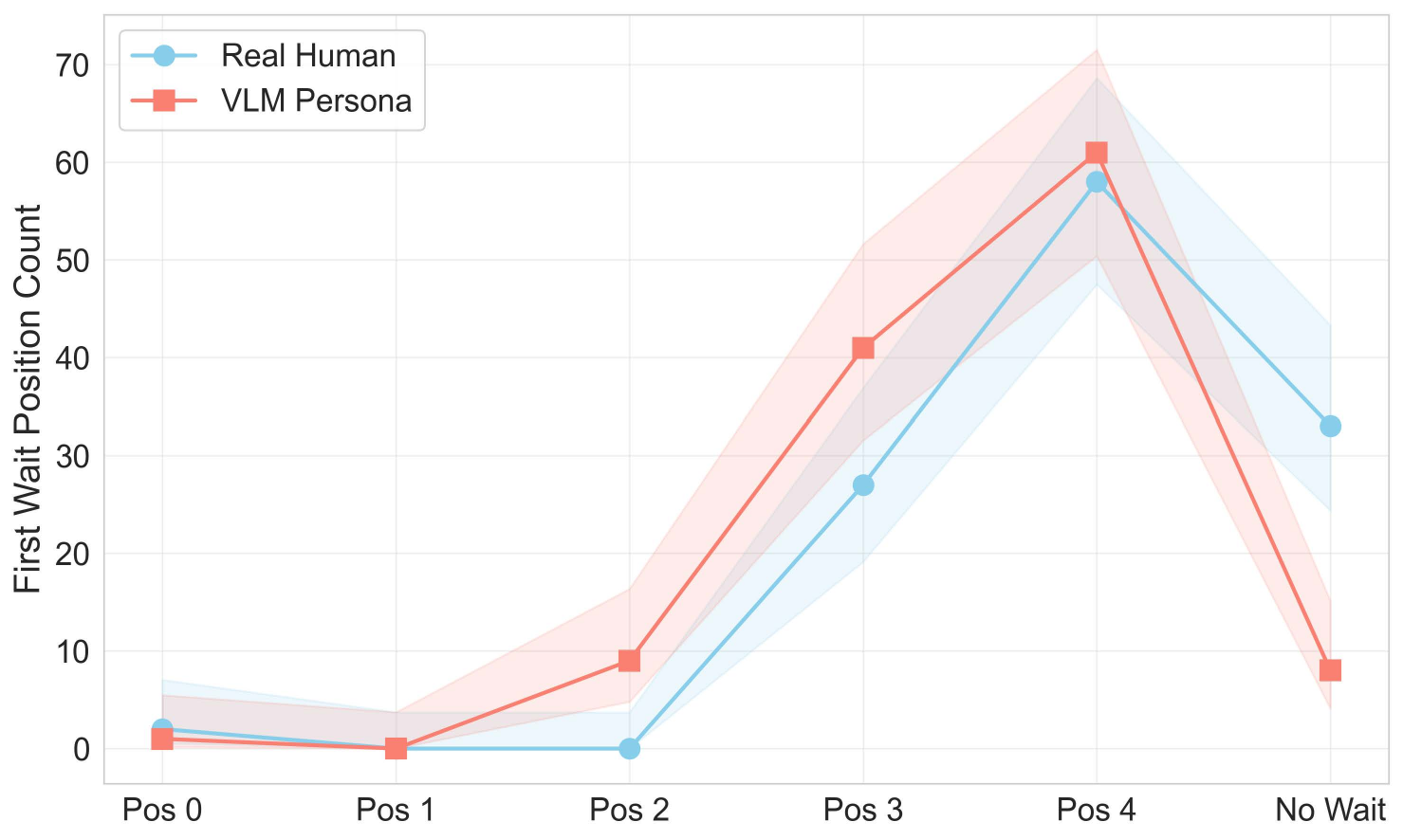}
        \caption{The cumulative count of the first wait (choose to stop) position with 95\% Wilson CI.}
    \end{subfigure}%
    \hspace{0.2cm} %
    \begin{subfigure}{0.45\textwidth}
        \centering
        \includegraphics[width=0.9\linewidth]{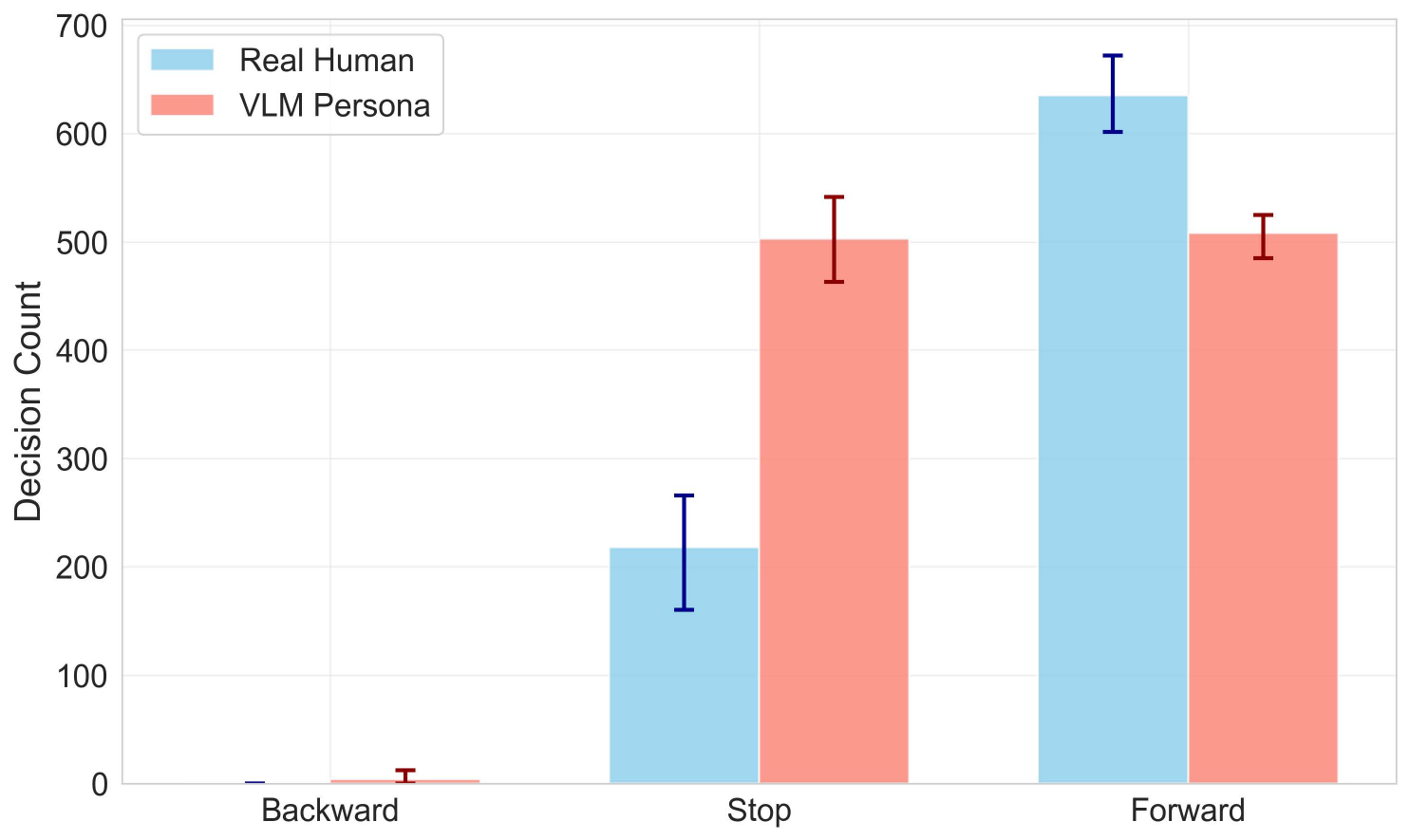}
        \caption{The cumulative count of decisions among three types with 95\% Wilson CI.}
    \end{subfigure}%
    
    \par\medskip

    \begin{subfigure}{0.9\textwidth}
        \centering
        \includegraphics[width=0.98\linewidth]{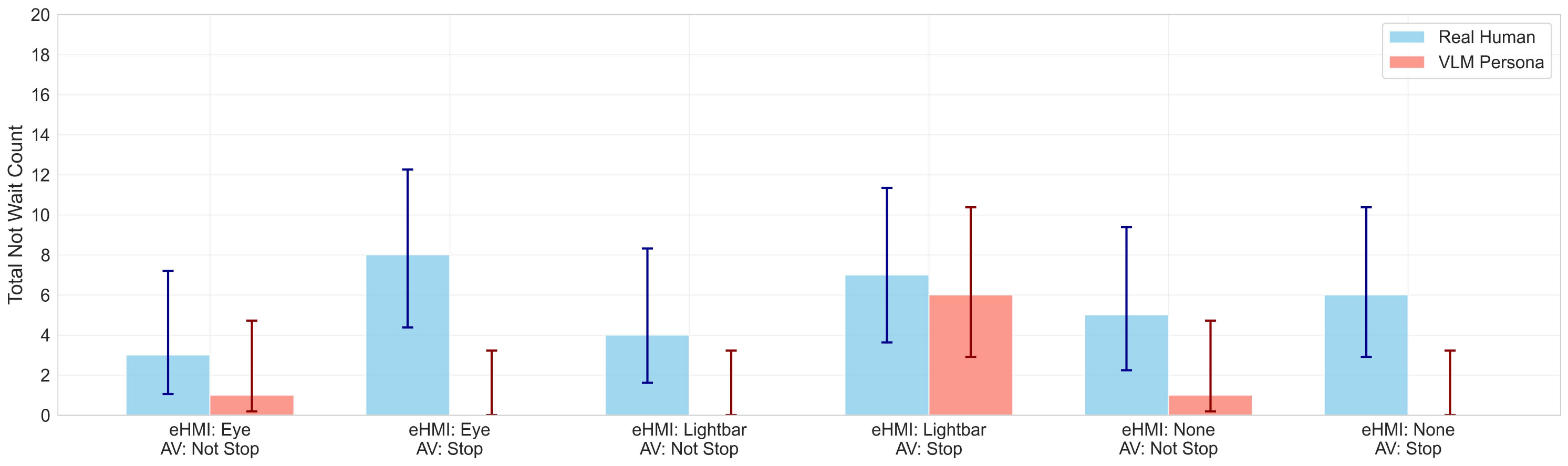}
        \caption{The cumulative count of "waiting behavior” (choose to stop) in 6 conditions with 95\% Wilson CI.}
    \end{subfigure}%
    
    \caption{Comparison of discretized trajectories between real humans and VLM personas, including the cumulative count of first-wait positions, decision-making across all experiments, and waiting tendencies across various scenarios.}
    \Description{Comparison of discretized trajectories between real humans and VLM personas, including the cumulative count of first-wait positions, decision-making across all experiments, and waiting tendencies across various scenarios.}
    \label{fig:new_2}
    \end{figure*}

The results of this comparison are shown in \autoref{fig:new_2}. As these figures are based on cumulative counts (y-axis) with 95\% CI, the findings represent behavioral tendencies. \autoref{fig:new_2} (a) shows the distribution of the first waiting position, which is where individuals choose the ``stop'' decision during the crossing. The two distributions are roughly aligned, increasing and peaking at position 4. This suggests that both real humans and VLM personas tend to walk, then stop near the road edge to observe the vehicle's movement and eHMI before making a final decision, showing VLMs are more conservative. In \autoref{fig:new_2} (b), we observed two differences in the distribution of decisions (forward, stop, backward). First, human participants never chose to move backward, but lowered their speed (slope change in the right part of \autoref{Fig:6}, which is hard to reflect in this discretizations figure. However, the VLM persona utilized this option. Second, humans showed a strong tendency to "move forward," while the VLM persona's decisions were more \textbf{evenly distributed} between ``forward'' and ``stop.''

\autoref{fig:new_2} (c) illustrates the total ``not wait'' count (keep waiting all the time) for all 20 individuals (either human or VLM), across the six experimental conditions. This analysis shows two tendencies. The first is the logical incoherence for VLM personas. For human participants, the ``not wait'' count was logically higher in the ``AV stop'' cases. However, the VLM persona showed a converse result in the Light-eHMI condition, tending to stop more during the crossing when the AV chose to stop. Second, we observed that humans often slow down (slope change in the trajectory curve) to observe while still moving, whereas the VLM treats this ``slow-down''  behavior as a hard  ``stop.'' We hypothesize this is affected by the definition of speed thresholds in the video simulator parameter setting.

\subsection{Subset Analysis: Effect of group size on similarity}

We clarify that all our comparisons are between-subjects. From a theoretical standpoint, an LLM is an aggregate model trained on population-level data \cite{liu2024evaluating, wang2025large}, indicating that a VLM persona is unlikely to capture the high-variance behavior of a single individual. Conversely, we hypothesize it should be highly effective at mimicking the central tendency of a group, where individual noise is averaged out.

\begin{figure*}[htbp]
\includegraphics[width=1\textwidth]{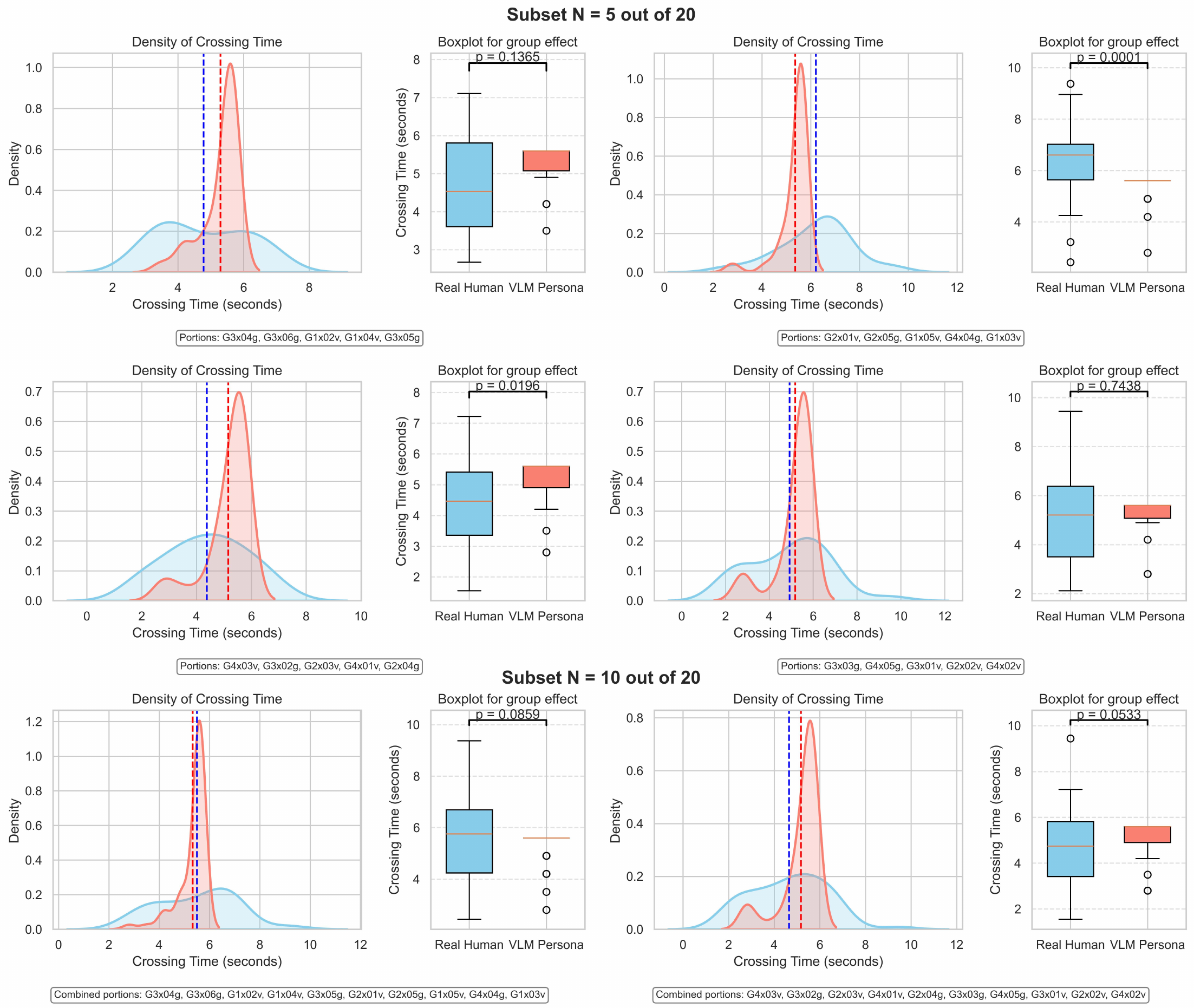}
\caption{The subset analyses comparing the VLM persona subset with human subsets: N=5 (first two rows, where 20 individuals are divided into four groups) and N=10 (third row, where 20 individuals are divided into two groups), based on crossing time.}
\Description{The subset analyses comparing the VLM persona subset with human subsets: N=5 (first two rows, where 20 individuals are divided into four groups) and N=10 (third row, where 20 individuals are divided into two groups), based on crossing time.}
\label{fig:new_4}
\end{figure*}

Our data empirically supports this. We conducted subset analyses by comparing the VLM persona cohort to human subsets of N=5 (\autoref{fig:new_4} (first two rows)), N=10 (\autoref{fig:new_4} (the third row)), and N=20 (\autoref{Fig:7}). We found that with small, high-variance samples (e.g., N=5), statistically significant differences between the human and VLM groups sometimes appeared. However, as the sample size increased, these differences diminished. The p-value increased, moving from statistical significance (p < 0.05 when N = 5 (top right and middle left group in \autoref{fig:new_4})) toward non-significance (p > 0.05 when N = 10, and P > 0.01 when N = 20).

This leads to our conclusion that the VLM's reliability to mimic human tendencies strengthens as the comparison group size increases. The VLM effectively captures the "average" human response, which only becomes statistically apparent when compared against a sufficiently large and stable human cohort. These can serve as the guidelines for other researchers using VLM personas for their simulation, that \textbf{1) VLM personas might be poor proxies for individuals or small groups. 2) VLM personas are better proxies for groups or populations for human-AV interaction.}

\section{Subjective Feeling Analysis: the comparison of subjective perception}

\subsection{Likert scale questionnaire rating}

\begin{figure*}[htbp]
\includegraphics[width=1\textwidth]{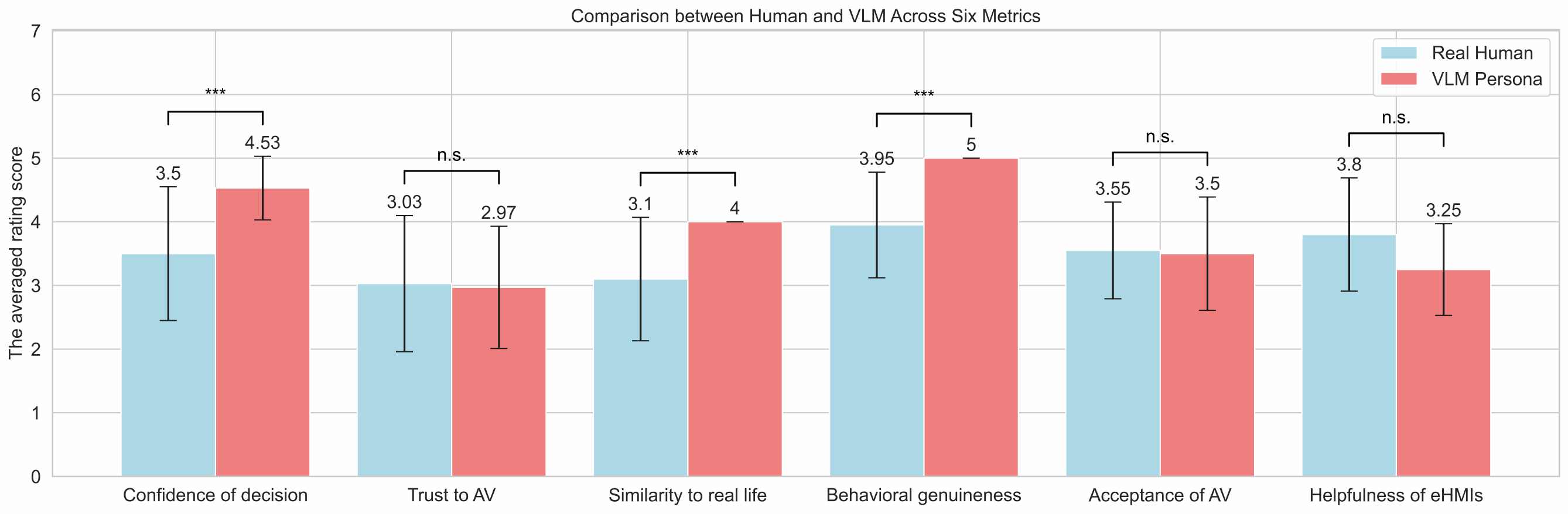}
\caption{Bar plot of average scores across six metrics, rated on a 5-point Likert scale by real humans and VLM personas.}
\Description{Bar plot of average scores across six metrics, rated on a 5-point Likert scale by real humans and VLM personas. These six metrics were adapted from three related research domains. The first two are commonly used in road-crossing scenario evaluations to assess pedestrian models when facing AVs \cite{rasouli2017understanding}. The next two assess the ecological validity of field studies \cite{hoggenmuller2021context}. The final two are derived from eHMI research to evaluate communicational efficiency and user acceptance \cite{gui2022going, dey2020taming}.}
\label{Fig:8}
\end{figure*}

The questionnaire analyses computed the effect of testing group (real human vs. VLM persona) on 5-point Likert ratings. We tested all six subjective rating metrics for normality and found non-normal distributions (\autoref{Fig:8}). Therefore, for the first two metrics (confidence and trust), rated 6 times during the study (n=240), we used a non-parametric two-factor ANOVA (group and condition). For the remaining four metrics (similarity to real life, behavioral genuineness, acceptance of AV, and helpfulness of eHMI) rated once after the study, we used unpaired Wilcoxon tests.

\autoref{Fig:8} shows significant differences in \textbf{confidence} (p<0.001), \textbf{similarity to real life} (p<0.001), and \textbf{behavioral genuineness} (p<0.001). The VLM group reported higher confidence (mean=4.53, SD=0.50) than the human group (mean=3.50, SD=1.05), higher similarity to real life (VLM: 4.00, SD=0.00; Human:3.10, SD=0.97), and higher behavioral genuineness (VLM: 5.00, SD=0.00; Human: 3.95, SD=0.83). 

No significant differences were found in \textbf{trust} (p=0.552), \textbf{acceptance of AV} (p=0.877), or \textbf{helpfulness of eHMI} (p=0.214). Trust ratings were highly similar (Human: 3.03, SD=1.07; VLM: 2.97, SD=0.96), as were acceptance (Human: 3.55, SD=0.76; VLM: 3.50, SD=0.89). Helpfulness also showed a non-significant trend with human group (Human: 3.80, SD=0.89; VLM: 3.25, SD=0.72).

\subsection{Thematic analysis for both human and VLM}
\label{interview}

Previous research has conducted various textual analyses to compare humans and LLMs, aspects such as response length, word usage, tone, writing style \cite{reinhart2025llms}, and human-likeness (e.g., through Turing tests \cite{jones2025large}), showing that LLMs can generate human-comparable text. In contrast, our study investigates whether VLMs are capable of reasoning with a similar depth and breadth as humans. That is, we explore whether VLMs focus on similar conceptual aspects, enabling us to extract key insights to support further AV-pedestrian interaction. To analyze the open-ended responses (from Q5 to Q7), we used thematic analysis \cite{braun2021can} and compared the themes (key findings) for the all three questions from both VLM-generated and human-generated responses.

\textbf{VLM responses:} One experimenter manually conducted a thematic analysis for all VLM responses. After coding and refining key terms, the experimenter then summarized four key findings from three aspects:

\vspace{-0.05cm}
\begin{itemize}
  \item \textbf{Interpretation of eHMI Meaning:}
    \begin{enumerate}
      \item The ``eye'' display indicates that the AV is aware of the pedestrian's presence.
      \item The ``light'' display follows traffic signal conventions (green = safe to cross; red = do not cross).
    \end{enumerate}

  \item \textbf{Potential Issues:}
    \begin{enumerate}
      \setcounter{enumi}{2} 
      \item Limited practical visibility of the light strip in strong sunlight conditions.
    \end{enumerate}

  \item \textbf{Preference for eHMI:}
    \begin{enumerate}
      \setcounter{enumi}{3} 
      \item eHMIs help reduce cognitive load and uncertainty for pedestrians.
    \end{enumerate}
\end{itemize}

\textbf{Human responses:} Another experimenter conducted a thematic analysis of field study responses without viewing the results from the VLM side. Six key findings were identified using the same thematic structure:

\vspace{-0.05cm}
\begin{itemize}
    \item \textbf{Interpretation of eHMI Meaning:}
    \begin{enumerate}
        \item Gaze direction is interpreted as a social cue indicating AV awareness.
        \item The light strip is intuitively understood via familiar traffic light metaphors.
    \end{enumerate}
    
    \item \textbf{Potential Issues:}
    \begin{enumerate}
        \setcounter{enumi}{2} 
        \item The strong sunlight leads to the absence of the light-based eHMI, causing uncertainty and confusion.
        \item Ambiguity in interpreting both ``eye'' and ``light'' displays.
        \label{eyeeHMI}

        For example, 2 participants interpret that ``\textit{I think the 'eye gaze' means AV's next driving direction }'' rather than the "awareness of pedestrians" as the other 18 participants' interpretation, which is the opposite meaning. Regarding the light, 3 participants mentioned it is hard to understand ``\textit{I am not sure whether the green indicates to AV or to me}''. 
    \end{enumerate}

    \item \textbf{Preference for eHMI:}
    \begin{enumerate}
        \setcounter{enumi}{4} 
        \item Relying only on vehicle speed and distance increases anxiety, and eHMIs provide reassurance.
        \item Three participants preferred the light strip as it felt more direct, whereas two preferred the eye display because it conveyed more emotional intent.
    \end{enumerate}
\end{itemize}

\vspace{-0.05cm}
\textbf{Comparative Insights:} The themes identified from the VLM responses (Findings 1-4) closely align with several core themes observed in the human data (Findings 1, 2, 3, and 5). However, the VLM responses lacked certain subtleties present in the human interpretations. They did not capture divergent interpretations expressed by a minority of participants (Finding 4), nor did they reflect user preferences related to eHMI choices (Finding 6). Finding 4 shows that while VLMs can replicate the consensus-level reasoning found in human analysis, they struggle to surface minority viewpoints. Finding 6 shows that VLMs provide direct answers to a given query, while human responses often incorporate wider contexts and associative thoughts that extend beyond the initial question. That is, although VLM-generated insights are directionally consistent with human interpretations, they offer narrower and shallower results.

\section{Interview with HCI researcher}

To answer the RQ2, we conducted interviews with five human-automation researchers whose work requires field studies. We chose the sample size N=5 to prioritize quality over quantity. We decided to conduct interviews after the experiments and data comparison to establish a necessary performance baseline for VLM persona simulation. Each interview lasted about one hour and included a concept introduction, a VLM simulation demonstration, a review of the comparison results, and a structured discussion with senior researchers. This aligns with qualitative principles \cite{nielsen1993mathematical}, where a small expert sample with deep engagement is sufficient to reach thematic saturation. 

Among the five interviewee, two are assistant professors focused on robotics, two are associate professors researching remote tele-operation and social interaction, and one is a senior researcher with over 10 years of experience in human-robot interaction. Our interview contains three parts. We first asked about the role of field studies in their research and the difficulties they encountered. Second, we introduced our concept, demonstrated the simulation, and showed comparison results. We then asked for their general impressions and whether they would consider using VLM personas to support their field studies. Finally, we asked how they might integrate VLM personas into their own research. 

\subsection{Field study importance and difficulties}

All five researchers emphasized the necessity of field studies. They mentioned two key reasons: \textbf{1) Understanding real-world problems}: Three interviewees mentioned that field studies help them better understand human behavior and test interactions in real-world settings. \textbf{2) Limitations of lab studies}: Two noted that lab-based experiments often lack authenticity, making it harder to evaluate natural human-robot interactions.

Regarding challenges they faced in their past research, we summarized four main categories: \textbf{1) Environmental constraints:} Finding suitable controllable locations is difficult. Unpredictable human behavior also adds complexity. \textbf{2) Participant recruitment:} It’s often hard to recruit the target population. Moreover, unexpected behavior by participants can pose safety risks. \textbf{3) Hardware and technical issues:} Their field studies frequently suffer from unreliable hardware, such as poorly integrated systems, connectivity problems or technical bottlenecks. \textbf{4) Resource demands:} Field studies demand more time, stricter safety standards, and stronger coordination than lab studies, requiring detailed planning and effective knowledge sharing.

\subsection{Willingness to use VLMs and key concerns}

All interviewees expressed interest in the potential of VLMs to simulate human behavior, noting that VLMs reflect how humans process information. We asked about their willingness to adopt VLM personas to support their field study. The result is 2 researchers said "yes", 2 said "no", and 1 remained undecided (see \autoref{fig:teaser}, red part on the right). They raised two major concerns. The first concern is \textbf{stability of VLM}. Several researchers worried about the consistency of VLM outputs. If VLM behavior is unreliable, it undermines the credibility of the simulation. Human behavior also contains randomness. Further research to explore how randomness in both real human and VLM personas (i.e., temperature \cite{pavlovic2024understanding}) influences the similarity in comparison is important. The second concern is \textbf{applicability to special populations}. While VLM personas may reasonably simulate the average person, they may not represent specific groups (e.g., children) accurately. Because children's stride speed and perception of the world are different. Meanwhile, proving their usability in these contexts is difficult since it is hard to pass the ethical review to call children in a field study for comparison.

\subsection{Potential Use Cases}

Even the researchers who previously answered "no" in a field study expressed openness to using VLMs as supplementary tools in other steps within the overall research process.

\textbf{1) Formative or Pilot study tool}: Three interviewees mentioned that using VLM personas as a preliminary study tool could be beneficial, as long as the results are \textbf{not over-interpreted}. As a formative study tool, ``\textit{this approach can provide an initial understanding of interactions and assist in designing future experiments with real human participants}''. As a pilot study tool, ``\textit{it helps improve simulations before conducting field studies, allowing researchers to be better prepared}''.

\textbf{2) Application for under-represented groups}: One interviewee suggested that the approach could be useful for studying hard-to-reach groups (e.g., individuals of extreme age \cite{hensch2021effects} or with disabilities). However, before applying the method in such contexts, the interviewees emphasize that ``\textit{it must be demonstrated that VLM personas can provide reliable simulations in the specific situation}''. At a minimum, a \textbf{clear user guideline} should outline all known limitations.

\textbf{3) Large-scale simulation}: Other interviewees emphasized the potential of using VLM personas to simulate a \textbf{large and diverse user base}. This could be valuable for preparing field studies by exploring a wide range of possible users. It could also support pretraining parts of systems that require behavioral data. However, both interviewees warned ``\textit{never blindly increasing the sample size in field studies by mixing real and synthetic data, as this can be difficult to justify}''.

\section{Discussion}

In this discussion, we first conclude our results into three actionable guidelines (Sec. \ref{guideline}) for researchers utilizing VLM personas in evaluation simulations, highlighting critical measures and potential pitfalls. We then integrate these guidelines with expert insights from five field researchers to propose a research process diagram (Sec. \ref{diagramafterinterview}). This diagram (\autoref{Fig:9}) describes how VLM personas can be effectively incorporated at various stages of the research workflow. Finally, we expand our findings from the Human-AV domain to broader embodied interaction (Sec. \ref{extension}). Although ``embodied'' in HCI usually refers to technologies physically integrated with the human body, in this project we use ``embodied'' specifically to describe the context of simulated personas. By analyzing the input-output modalities within embodiment literature, we identify the necessary technical steps to expand VLM persona simulations to broader domains.

\subsection{Usage guidelines from the comparison result}
\label{guideline}

Our behavioral analysis indicates that while VLM personas successfully replicate the central tendency of human actions (showing no statistically significant differences in mean and median), they fail to capture the full spectrum of human variability. The VLM data exhibits significantly lower standard deviations, effectively representing only a ``safe'' subset of the broader human behavioral distribution. Furthermore, our subset analysis demonstrates that small sample sizes exacerbate sampling bias. \textbf{Guideline 1: VLMs are reliable for predicting general behavioral trends, but not for modeling extreme or highly variable human actions. A sufficient volume of personas is required to mitigate sampling bias and stabilize the distribution.}

A distinct pattern emerged in the questionnaire data depending on the reference point. When VLM personas assessed intrinsic states (e.g., self-confidence, genuineness), they exhibited the ``hyper-confidence'' phenomenon often observed in LLMs \cite{sun2025large, davidson2024self}, leading to inflated ratings. Conversely, when evaluations were directed outward at external objects (e.g., trust in the AV, eHMI utility), the VLM ratings aligned closely with human ground truth. \textbf{Guideline 2: Avoid asking VLM personas to rate their own internal states. To minimize calibration errors, framing should focus on evaluating external objects or system features.}

In the interview simulation by VLM persona shows that the VLM acts as a ``consensus engine,'' producing logical but homogeneous text responses. It failed to capture the minority perspectives and unexpected interpretations found in our human dataset. \textbf{Guideline 3: Use VLM personas to summarize standard user expectations, but do not rely on them to uncover overlooked usability issues. Human participants remain essential for capturing diverse, non-standard, or creative interpretations.}

\subsection{Cross discussion between comparison result and HCI researcher interview} 
\label{diagramafterinterview}

\begin{figure*}[htbp]
\includegraphics[width=1\textwidth]{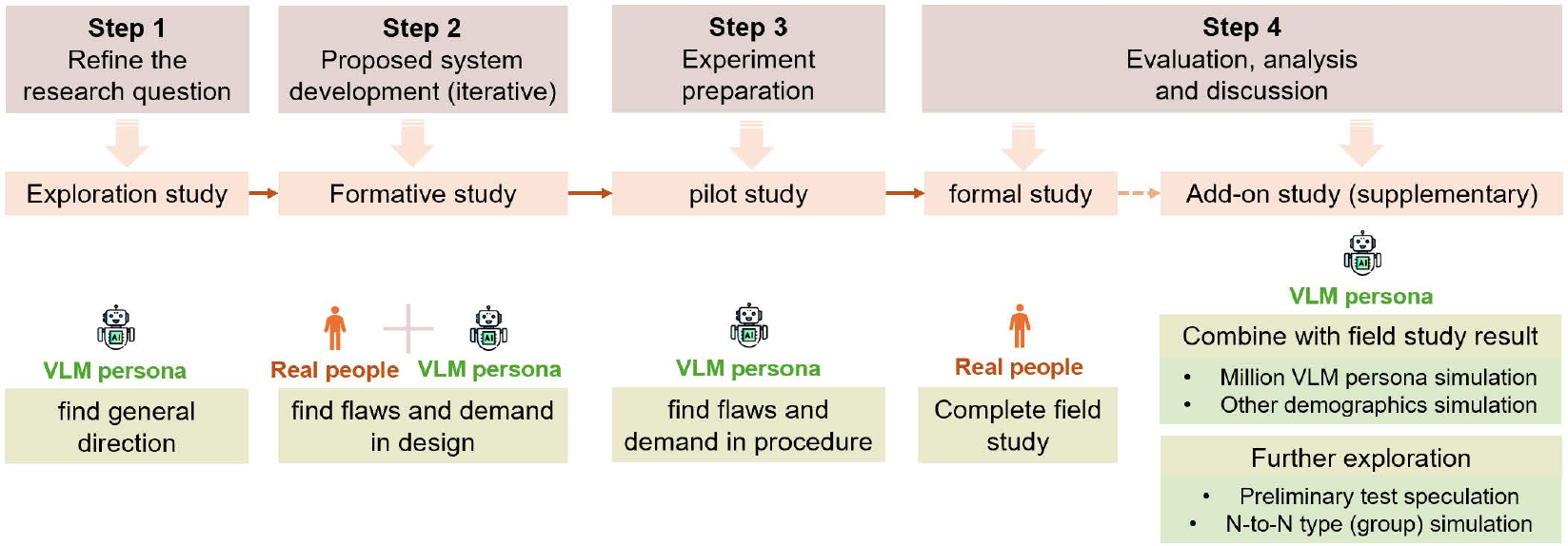}
\caption{Based on the comparison of the two parallel experiments and insights from HCI researchers’ envisioned use cases, we created a research process diagram highlighting how VLM personas can be incorporated at various stages of the HCI research workflow.}
\Description{Based on the comparison of the two parallel experiments and insights from HCI researchers’ envisioned use cases, we created a research process diagram highlighting how VLM personas can be incorporated at various stages of the HCI research workflow.}
\label{Fig:9}
\end{figure*}

The typical research process is generally divided into four stages \cite{lazar2017research}: initial exploratory research, iterative system design, pilot study, and formal field study. We integrate guidelines from our comparative analysis with use cases from follow-up interviews to illustrate how VLM personas can support each stage (\autoref{Fig:9}).

In the exploration stage (\autoref{Fig:9}, column 1), VLMs have shown the capacity to generate general insights, allowing researchers to quickly form a preliminary direction in unfamiliar domains. VLM's ability to synthesize widely accepted knowledge can help accelerate early-stage ideation.

During the formative study stage (\autoref{Fig:9}, column 2), VLM personas show potential but have key limitations. They struggle to capture ambiguous or minority perspectives (Guideline 3) and to reflect first-person perceptions (Guideline 2). Thus, while feasible at this stage \cite{o2018charrette, miaskiewicz2011personas}, real human feedback remains necessary to ensure adequate coverage and depth.

In the pilot study stage (\autoref{Fig:9}, column 3), VLM personas are useful for preliminary testing. They can simulate general behavior to identify procedural flaws, assess clarity of instructions, and ensure the study setup aligns with user expectations. This pre-testing improves efficiency before involving human participants.

In the field study stage (\autoref{Fig:9}, column 4), real human data remains irreplaceable. However, VLM personas can supplement studies (\autoref{Fig:9}, column 5) by expanding participant diversity, provided their alignment with real data is carefully validated. They may also support preliminary hypothesis testing or scenario exploration to inform subsequent empirical research.

\subsection{Extension to Broader Embodied Interaction}
\label{extension}

Our method can be extended beyond the specific case of AV-human interaction to the wider domain of embodied interaction \cite{fung2025embodied, duan2022survey}. Our primary focus on AVs was driven by necessity (the high costs and safety risks associated with field studies in traffic environments). The proposed evaluation framework with VLM personas offers significant value for other domains where distinct constraints exist. For example, even in scenarios with lower physical risk, such as indoor human-robot interaction (e.g., household robot \cite{shiokawa2025beyond}), evaluating with real human participants presents challenges. These include recruitment difficulties for underrepresented populations (e.g., individuals with disabilities) \cite{banas2019recruiting}, designing a user study to overcome VR headsets' uncomfortableness and motion sickness \cite{koohestani2019knowledge, chattha2020motion}, and the logistical complexity of analyzing cognitive states in multi-party coordination tasks (e.g., disaster response) \cite{chew2024human, kantharaju2021social}.
 
To generalize the VLM persona method for these broader embodied tasks, we believe the future research needs to address two main technical categories \cite{wilson2003embodying}: input modalities \cite{vatavu2024ai} and output fidelity \cite{grubert2015multifi}. For example, in our field study of AV-pedestrian interaction, the LLM takes vision input and generates human mimicry behaviors. \textbf{Regarding the input modalities, we call the multimodal input research to explore pipelines that enable VLMs to acquire information through diverse senses, mimicking human perception.} In this study, we processed visual inputs by segmenting 30 videos into 150 clips, allowing the VLM to build a unique ``interaction memory'' based on step-by-step decisions. However, broader embodiment requires extending this beyond vision. Future simulators may integrate audio \cite{tao2025audio}, haptic \cite{wang2020miniature}, olfactory \cite{maggioni2020smell}, or even direct VR visual feeds \cite{zhong2025unrealzoo}, effectively giving the VLM the sensory capacity to perceive specific physical environments just as a human participant would. \textbf{Regarding the output fidelity, we call for research to validate the human-likeness of the VLM's generated data across different dimensions.} For example, in our study, we compared generated behavior (i.e., locomotion \cite{wang2024grounding}) and subjective perceptions toward AV. However, embodied interaction involves complex physiological and psychological states. Future work should investigate the VLM's ability to simulate: emotional responses (e.g., Ekman's model \cite{ekman1999basic}), cognitive load (e.g., NASA-TLX scores \cite{hart1988development}), and physiological signals (e.g., heart rate, EEG, ECG patterns) \cite{chiossi2024physiochi, chiossi2023exploring, chiossi2024optimizing}. Validating these outputs would require new comparative studies between human physiological data and VLM persona generated results.

Besides technical feasibility, we need the application-oriented exploration to identify the specific domains where VLM personas could uniquely benefit embodied interaction research. Here are three example applications. The first is smart city robotics. Unlike AVs on roadways, urban service robots often navigate sidewalks and shared public spaces. VLM simulations could safely test these frequent and socially complex interactions. The second is augmented reality gaming. Applications like \textit{Pokémon Go} require physical movement and spatial interaction that are difficult to replicate in static lab settings. VLM personas could simulate user behavior in these unbounded real-world environments. The third is a high-risk cognitive investigation. In scenarios such as disaster preparedness, real-world testing is both dangerous and ethically fraught. VLM personas offer a method to study embodied cognition and adaptive decision-making under pressure without risking human lives. These studies require participants to engage in real-time spatial sensing and behavioral and emotional responses. They are typically large-scale, focus on societal rather than individual systems, and rely on extensive physical infrastructure, real-time sensing technologies, and a significant number of participants. The demonstrated similarity between VLM and human responses in certain aspects validates the use of VLM personas for supporting the research. For example, it streamlines formative and pilot studies by minimizing reliance on human participants without compromising the study quality.

\subsection{Limitation and Future work}

This study used GPT-4.1. That is, the findings and guidelines may evolve as model capabilities improve. For example, unlike humans, who rely on binocular vision for depth perception, the current VLM lacks this ability. Future advancements in binocular or multi-view perception could enhance performance on tasks requiring spatial reasoning. Opportunities remain to improve simulated evaluation. While this study used real-world video stimuli, it is unclear whether alternative visual formats (e.g., VR models or animations) would yield comparable results. If validated, VLM personas could further simplify the pipeline by reducing the need for on-site filming.

One important limitation is the prompt of the construction VLM persona. \refine{Our extraction process relies entirely on LLM reasoning driven by the designed prompts, making the modeling result highly rely on the model performance. While feasibility can be examined through simulation reasoning traces, the overall interpretability remains limited. An important direction for future work is to develop more principled and transparent modeling approaches that improve both variability and interpretability. Building on our findings, future work could analyze reasoning traces (Appendix B) to inform parameter-weighted model structures or to refine prompt engineering (e.g., by modifying Listing 1).}

Constructing robust AI personas remains an unresolved challenge in NLP due to two main issues. First, there is no standardized design framework or consensus on an ``optimal'' methodology \cite{tseng2024two, kim2024persona, luz2025helpful}. Second, robust benchmarks are lacking to validate whether a persona faithfully replicates a target profile across dimensions \cite{zhou2025personaeval}. Prior work shows that persona-shaping prompts cannot fully override models' internal preferences \cite{kim2024persona}. Although alignment with general human values and social norms is essential for safety, it may reduce output diversity and constrain behaviors that conflict with safety policies \cite{openai_bot_impersonating_presidential_candidate_2024, openai_using_chatgpt_agent_policies_2025}. This limitation can affect the diversity and fidelity of LLMs in role-playing tasks such as those in our study.

\section{Conclusion}

VLM personas can partially simulate human behavior in embodied tasks. Our study extends this line of work by exploring how this capability can support field studies in HCI research. To figure out the quality of using VLM personas as a tool (either in pilot studies or to augment field studies), we \textbf{conducted two parallel experiments}: 1) a field study with human participants and 2) a video study using VLM personas, enabling both quantitative and qualitative comparison. The findings reveal no significant overall differences, although there is reduced variability and diversity.

We identified \textbf{three key guidelines} that outline the aspects under which VLM persona responses align with those of real humans. \textbf{1) Use VLM-generated data cautiously}, as it can reflect human responses but only captures a subset due to reduced variability. \textbf{2) Frame evaluations in third-person}, since VLMs tend to exaggerate when responding from a first-person perspective. \textbf{3) Match VLM use to research goals} as they seem suitable for broad exploration of new topics, but less effective for uncovering nuanced or overlooked issues in familiar domains. To explore broader applicability, we interviewed HCI researchers and developed a \textbf{research process diagram} showing how VLM personas can be integrated into the HCI workflow.

\section*{Open Science}
The code and prompts for replicating the VLM-simulated field study task are available at: \url{https://github.com/ApisXia/PersonaVLM}.

\begin{acks}
This work was supported by JST CRONOS, Grant Number JPMJCS24K8, Japan. Mark Colley was supported by a Canon Research Fellowship.
\end{acks}

\bibliographystyle{ACM-Reference-Format}
\bibliography{VLM}

\clearpage
\appendix
\section*{Appendix}

\section{Persona example}
\label{persona}
\begin{lstlisting}[style=promptstyleAppendix]
"test16": {
    "name": "G3x03g",
    "description": "Here is the basic demographics of the people: The age is 21. The gender is male. A Chinese people currently lives in Japan for 3 years. Currently is pursuing the bachelor degree.\n\nRegarding the possible behavior when crossing a street: This individual may experience heightened awareness or stress in ambiguous situations, making them more reactive to traffic or signals. However, their outgoing nature might lead them to follow others quickly once they perceive a window of opportunity. Driven and imaginative, they might respond creatively in unexpected scenarios, though they may act more for personal gain than shared safety.",
    "decision_criteria": [
      "Impression of autonomous driving: I feel that autonomous driving has a lot of potential. I've never been in one myself, but I'd be willing to give it a try.",
      "Use case of autonomous driving: Maybe I will choose autonomous driving as my main form of daily transportation",
      "Emotion of autonomous driving: I would probably feel curious and excited.",
      "Concern of autonomous driving: safety",
      "Expectation of autonomous driving: I would want them to have a greater ability to communicate externally, for instance, an indicator light to alert others that the vehicle is in self-driving mode."
    ]
  },

  "test03": {
    "name": "G2x02v",
    "description": "Here is the basic demographics of the people: The age is 46. The gender is male. A Japanese people currently lives in Japan. A Business Owner with PhD degree.\n\nThis individual handles stress well and remains composed even in tense environments. He values his privacy and typically keeps to himself or interacts with a tight-knit group. Curious and mentally agile, he enjoys exploring new ideas and unconventional solutions. He cares about the wellbeing of others and acts with integrity. Organized and reliable, he sets goals and sticks to them. When crossing the street, he is likely to observe carefully, make a calculated decision, and proceed steadily, ensuring safety without over-relying on others.",
    "decision_criteria": [
      "Impression of autonomous driving: I rode a automated Uber in the USA and felt positive towards that. I am hoping that most of the cars have autonomous driving feature to reduce the human error originated traffic accident.",
      "Use case of autonomous driving: Yes I can. My car has lane keep assisting feature using the camera or lidar. I am strongly hope to have automated vehicle soon.",
      "Emotion of autonomous driving: Happy and it will reduce our burden driving long distance.",
      "Concern of autonomous driving: Software and hardware malfunction which leads to the traffic accident",
      "Expectation of autonomous driving: Safety"
    ]
  },

  "test14": {
    "name": "G4x03v",
    "description": "Here is the basic demographics of the people: The age is 23. The gender is male. A Algeria people currently lives in Japan for 2 months. Currently is pursuing the master degree.\n\nRegarding the possible behavior when crossing a street: Sociable and adaptive, this person may follow others across if the group leads, but with a sense of calm confidence. Their curiosity may lead them to observe and interpret different crossing scenarios, while their structured mindset ensures they do not act impulsively. Cooperative and emotionally stable, they will likely take a balanced and mindful approach to street-crossing.",
    "decision_criteria": [
      "Impression of autonomous driving: I have never been in a self driving vehicul , but from what I have seen , it looks like a really pronising technology and I beielive it will be the main way of transportation in the future.",
      "Use case of autonomous driving: I would use automated vehilcels for getting around if developed enough to a point where it can be safely completly autonomous.",
      "Emotion of autonomous driving: happy and hopefull",
      "Concern of autonomous driving: there is the fear if it malfunctioning which can be quite dangerous but with enough precautions that should never happen",
      "Expectation of autonomous driving: just it's ability to be completly safely autonomous"
    ]
  },

  "test15": {
    "name": "G3x02g",
    "description": "Here is the basic demographics of the people: The age is 27. The gender is female. A Chinese people currently lives in Japan for 5 years. Mostly get the master degree and become a PhD student later.\n\nRegarding the possible behavior when crossing a street: While composed and self-assured, this person may not prioritize others' safety unless directly affected. Their independent thinking might make them less reliant on conventional cues. They are likely to make fast decisions if they feel confident, but their approach remains calculated and based on personal logic rather than group consensus. Their motivation is often driven by efficiency and control.",
    "decision_criteria": [
      "Impression of autonomous driving: Risky, and not that reliable.",
      "Use case of autonomous driving: If I have to use the auto vehicle, I prefer the one that can assist human driving, 51% decision by human intelligence.",
      "Emotion of autonomous driving: Worried",
      "Concern of autonomous driving: If it gets out of control in a dynamic environment.",
      "Expectation of autonomous driving: If it can negotiate with me before important decisions when driving."
    ]
  },

  "test12": {
    "name": "G4x01v",
    "description": "Here is the basic demographics of the people: The age is 24. The gender is male. A Indian people currently lives in Japan for 2 years. Currently is pursuing the master degree of engineering.\n\nRegarding the possible behavior when crossing a street: Calm and observant, this individual would approach crossing with balanced awareness. While confident and socially aware, they are unlikely to take unnecessary risks. Open to new experiences, they may explore different options but still follow rules and respect shared safety. Their cooperative nature makes them more likely to follow pedestrian signals or group movements while maintaining awareness of their surroundings.",
    "decision_criteria": [
      "Impression of autonomous driving: A good invention for multitasking on the go.",
      "Use case of autonomous driving: I imagine using it while commuting.",
      "Emotion of autonomous driving: There are moments when the road's instability causes worry.",
      "Concern of autonomous driving: Nope",
      "Expectation of autonomous driving: More robust system"
    ]
  },

\end{lstlisting}

\section{Example of a fully simulated crossing task}
\label{cross}

\begin{lstlisting}[style=promptstyleAppendix]
Decision: forward
Reason: Based on my personality, I tend to feel some excitement and curiosity around autonomous vehicles, and I'm open to trying new things like checking how these vehicles react to pedestrians. The approaching car doesn't show strong external signals like specific eHMI light strips indicating its intention, which trips my expectation for better communication but doesn't deter me from taking action. Because I value my own curiosity and want to get a closer look (maybe even test if the AV will notice me waiting at the edge), I will move forward one step toward the road to observe its reaction more closely. However, I do remain cautious because safety is my primary concern with autonomous vehicles.
Confidence: 4/5 - I'm confident because the car is not close enough to be immediately dangerous and slowing/stopping soon, but I do want to proceed step-by-step.
Trust: 3/5 - I feel neutral trust due to lack of visible eHMI signaling. I'd like clearer communication from the AV to feel entirely assured.
New status: o-*-o-o-o-|ROAD

--- Time Step 1 (1.0s) ---
Current position: 1
Status: o-*-o-o-o-|ROAD
Analyzing video: data/250722_real_sim/no-ehmi_stop/split/pos1_time1.mp4
Extracted 24 frames from 24 total frames
Decision: forward
Reason: The car continues coming closer at a moderate speed, and I don't see any obvious external communication cues such as eHMI signals. Based on my curiosity and excitement about autonomous vehicles, I'm motivated to get even closer to see if its behavior changes or if any indicator will appear signaling its intentions to yield. Since safety is my main concern, I'm not stepping onto the road yet, but my drive to test the AV's response and my openness to the experience push me to move one more step forward. I'm attentive, looking for any sign of communication from the vehicle, as I expect autonomous cars to signal more clearly.
Confidence: 4/5 - I'm fairly confident my actions align with my personality - curious, outgoing, willing to test boundaries but not risk personal safety by entering the road just yet.
Trust: 3/5 - I feel somewhat neutral but optimistic about the AV. I'd like to see more active external communication before putting full trust in it, especially nearing the crosswalk.
New status: o-o-*-o-o-|ROAD

--- Time Step 2 (2.0s) ---
Current position: 2
Status: o-o-*-o-o-|ROAD
Analyzing video: data/250722_real_sim/no-ehmi_stop/split/pos2_time2.mp4
Extracted 24 frames from 24 total frames
Decision: forward
Reason: The autonomous vehicle is continuing to approach, but based on my heightened curiosity and excitement about seeing how it reacts, I'm motivated to step even closer to the road to test whether it will clearly communicate its intentions or slow down. My safety concern is always there, but my curious and driven side wants to provoke some reaction in the AV, especially since I expect external communication (like eHMI lights) but don't see any yet. Moving forward lets me observe its behavior up close while still not fully committing to the road, maintaining a balance between my bold and reactive impulses and my respect for personal safety.
Confidence: 4/5 - I'm quite confident in approaching further since I'm not actually crossing yet, and the distance still gives me room to pause and assess. My curiosity and excitement drive my action, but I retain a level of caution.
Trust: 3/5 - My trust in the autonomous vehicle is neutral. I believe it has potential and am willing to give it a try, but the lack of visible external signals means it hasn't quite met my expectations for communication. I don't fully trust it to stop for me until I see more proof.
New status: o-o-o-*-o-|ROAD

--- Time Step 3 (3.0s) ---
Current position: 3
Status: o-o-o-*-o-|ROAD
Analyzing video: data/250722_real_sim/no-ehmi_stop/split/pos3_time3.mp4
Extracted 24 frames from 24 total frames
Decision: stop
Reason: As I get to the edge before the road, my awareness and caution increase because the autonomous vehicle is still evidently approaching at a steady speed and has not shown any clear sign of slowing down or yielded - no eHMI or external indicator visible. My curiosity makes me eager to observe but since my main safety concern with autonomous driving is a lack of overt communication, and I'm not seeing anything to indicate that the vehicle is recognizing me or intending to stop, my reactive and aware side outweighs my excitement. I will stop at the edge, waiting to see if the AV finally reacts or gives me any indication it's safe to cross. My expectation for more transparency from AV is not met yet.
Confidence: 4/5 - I'm fairly certain I should stop here because the vehicle is too close and non-communicative, so I can't safely anticipate its actions yet.
Trust: 2/5 - My trust is low right now as the AV has not displayed any external communication or yielding behavior, which is central to my safety expectations.
New status: o-o-o-*-o-|ROAD

--- Time Step 4 (4.0s) ---
Current position: 3
Status: o-o-o-*-o-|ROAD
Analyzing video: data/250722_real_sim/no-ehmi_stop/split/pos3_time4.mp4
Extracted 24 frames from 24 total frames
Decision: stop
Reason: Based on my driven curiosity, I move forward quickly toward situations that capture my interest, especially when testing a new technology like autonomous vehicles. However, right now the vehicle is visibly approaching at a steady speed and distance, not slowing down. There's still no eHMI or indicator lights communicating that it sees me or intends to stop, which is a key expectation for me. My confidence comes from my habit in ambiguous street crossing situations of being hyper-aware of risk, and my priority is my own safety, especially as my main concern with AV is unpredictable behavior when they don't communicate. While I'm excited to try AVs, when facing a complete lack of communication and no signs of yielding, my stress/alertness is heightened. So, my instinct is to stop and watch the AV rather than step into the road. My decision is influenced by both my curiosity and my critical stance waiting for real external signals from the AV.
Confidence: 5/5 - I clearly see the vehicle has no yielding behavior or external signals; my safety-first trait in ambiguous moments makes stopping the only logical option.
Trust: 2/5 - The AV isn't communicating at all, missing my expectation for a visible eHMI or external indicator. That keeps my trust level low, even though I'm still open-minded.
New status: o-o-o-*-o-|ROAD

--- Time Step 5 (5.0s) ---
Current position: 3
Status: o-o-o-*-o-|ROAD
Analyzing video: data/250722_real_sim/no-ehmi_stop/split/pos3_time5.mp4
Extracted 24 frames from 24 total frames
Decision: stop
Reason: The autonomous vehicle is approaching very close now and its speed hasn't decreased throughout the frames, which makes me feel tense and hyper-aware of the possible danger. Even though I'm excited and curious about how AVs react to pedestrians, my main concern is safety, especially since there are still no external signals or eHMI to show the car has detected or intends to yield for me. My expectation is for clear communication, so the lack of any visible indicator directly affects my willingness to test this car further. My personal interest might normally drive me to challenge the situation, but since there is zero explicit sign from the car and it's very close, I'm not willing to risk it just for curiosity. My cautious side takes over under this uncertainty.
Confidence: 5/5 - The vehicle is extremely close and still moving, so my safety-based intuition and close observation are fully triggered, making me very certain I should wait.
Trust: 2/5 - The lack of reaction or any eHMI signal from the vehicle really undermines my trust in this AV, especially since my major expectation is obvious pedestrian communication.
New status: o-o-o-*-o-|ROAD

--- Time Step 6 (6.0s) ---
Current position: 3
Status: o-o-o-*-o-|ROAD
Analyzing video: data/250722_real_sim/no-ehmi_stop/split/pos3_time6.mp4
Extracted 24 frames from 24 total frames
Decision: stop
Reason: At this very close distance, the autonomous vehicle is continuing to approach steadily without showing any slowdown or visible eHMI external communication. My excitement and curiosity about AVs have definitely made me push forward in prior steps because I want to see how the technology interprets pedestrian intentions. However, safety is my central concern, and not receiving any explicit indicator or clear sign that the AV acknowledged me puts me in a high alert, slightly stressful mindset. Culturally, in Japan, I'm used to waiting for clear permission before crossing busy roads. Even as an outgoing and creative person, the personal risk here is too high without communication from the AV, so I must stop until I see strong evidence that it's yielded or intends to stop. My driven curiosity is overpowered by uncertainty and the lack of transparent vehicle signals.
Confidence: 5/5 - The car is extremely close, its approach speed is consistent, and there are still zero communication cues or slowdown. I feel absolutely certain it's unsafe.
Trust: 2/5 - My expectation is for AVs to communicate clearly when yielding; this one failed to show any signals or eHMI. My trust is currently low because that key expectation is not being met.
New status: o-o-o-*-o-|ROAD

--- Time Step 7 (7.0s) ---
Current position: 3
Status: o-o-o-*-o-|ROAD
Analyzing video: data/250722_real_sim/no-ehmi_stop/split/pos3_time7.mp4
Extracted 24 frames from 24 total frames
Decision: stop
Reason: Looking closely at the video, the autonomous vehicle is approaching steadily and is now extremely close, with no sign of slowing or stopping. Even though I'm an outgoing person who gets excited about new tech and enjoys testing boundaries, my main concern with AVs is always safety - especially when there's a lack of clear, external communication, such as eHMI or lightbars. Since I expect clearer signals (and none appeared) and my intuition is heightened now due to stress in this ambiguous moment, I'm staying put at the edge of the road. My concern over not being acknowledged outweighs my curiosity, and I'm definitely not stepping into the road.
Confidence: 5/5 - The vehicle's distance and speed leave no ambiguity. No eHMI or yielding signal confirms it's unsafe to proceed based on my expectations and personality.
Trust: 2/5 - My trust is low because my key expectation - explicit AV communication - is still unmet despite my repeated presence. This reinforces my doubt about how AVs handle pedestrian detection and interaction.
New status: o-o-o-*-o-|ROAD

--- Time Step 8 (8.0s) ---
Current position: 3
Status: o-o-o-*-o-|ROAD
Analyzing video: data/250722_real_sim/no-ehmi_stop/split/pos3_time8.mp4
Extracted 24 frames from 24 total frames
Decision: stop
Reason: Even though I'm highly curious and motivated to try new things like interacting with autonomous vehicles, and my personality may push me forward if there's even a small positive signal, my main concern is safety, especially in ambiguous or high-stress moments. The car is right in front of me, clearly did not show any eHMI or indicator it saw me or will stop. My expectation for direct, external communication before attempting to cross is not met at all. As a result, my heightened awareness and stress win out over my outgoing side - I will not move onto the road and choose to stop.
Confidence: 5/5 - The evidence is extremely clear: vehicle is very close, did not stop, and provided no signals. My behavior in these conditions is decisive.
Trust: 2/5 - My trust drops when AVs do not communicate or indicate intention. Lack of eHMI feeds my main worry and leaves me reluctant to rely on the vehicle.
New status: o-o-o-*-o-|ROAD

Simulation log saved to: /PersonaVLM/G3x03g/5_no-ehmi_stop/simulation_log.json

=== All Position Status Summary ===
Time 0: o-*-o-o-o-|ROAD (moved from 0 to 1 - forward)
Time 1: o-o-*-o-o-|ROAD (moved from 1 to 2 - forward)
Time 2: o-o-o-*-o-|ROAD (moved from 2 to 3 - forward)
Time 3: o-o-o-*-o-|ROAD (moved from 3 to 3 - stop)
Time 4: o-o-o-*-o-|ROAD (moved from 3 to 3 - stop)
Time 5: o-o-o-*-o-|ROAD (moved from 3 to 3 - stop)
Time 6: o-o-o-*-o-|ROAD (moved from 3 to 3 - stop)
Time 7: o-o-o-*-o-|ROAD (moved from 3 to 3 - stop)
Time 8: o-o-o-*-o-|ROAD (moved from 3 to 3 - stop)

Combining all video clips...

Output saved to: /PersonaVLM/G3x03g/5_no-ehmi_stop
Step videos saved to: /PersonaVLM/G3x03g/5_no-ehmi_stop/step_views
Combined video saved to: /PersonaVLM/G3x03g/5_no-ehmi_stop/all_agent_see.mp4

Conducting post-simulation questionnaire...
Extracted 216 frames from 216 total frames

\end{lstlisting}

\appendix

\end{document}